\pdfoutput=1

\documentclass[prd,aps,twocolumn,preprintnumbers,amssymb,nobibnotes,nofootinbib,longbibliography,superscriptaddress,10pt]{revtex4-2}

\usepackage[left=1in, right=1in, top=1in, bottom=1in]{geometry}   

\usepackage[utf8]{inputenc}
\usepackage[english]{babel}  
\usepackage{hyperref}
\hypersetup{
    colorlinks=true,
    allcolors={blue!60!black}
}

\usepackage{bbold}
\usepackage{graphicx,epsf}
\usepackage{amsmath,amsfonts,amssymb,amsbsy,mathrsfs}

\usepackage{tikz}
\usetikzlibrary{decorations.pathmorphing}
\usetikzlibrary{decorations.markings}
\usetikzlibrary{positioning, shapes, arrows}
\usetikzlibrary{calc}
\usepackage[compat=1.0.0]{tikz-feynman}

\tikzset{
quark/.style={postaction={decorate}, decoration={markings}},
scalar/.style={dashed,postaction={decorate}, decoration={markings,mark=at position .5 with {\arrow[#1]{latex}}}},
gluon/.style={decorate,decoration={coil,amplitude=3pt, segment length=4.7pt, pre length=.01cm, post length=.01cm}},
gluont/.style={decorate,decoration={coil,amplitude=3pt, segment length=3.50pt, pre length=.01cm, post length=.01cm}},
}

\usepackage{caption}
\usepackage{subcaption}
\usepackage{physics}

\usepackage{color,xcolor}

\usepackage{array,multirow}
\usepackage{stackrel}

\usepackage{ifdraft}

\usepackage{placeins}
\usepackage{slashed,shuffle} 
\usepackage{adjustbox}

\usepackage{comment}
\usepackage[normalem]{ulem}

\usepackage{slashed}

\usepackage{cleveref}
\usepackage{pifont}
\newcommand{\cmark}{\ding{51}}%
\newcommand{\xmark}{\ding{55}}%

\DeclareUnicodeCharacter{27E8}{{\langle}}
\DeclareUnicodeCharacter{27E9}{{\rangle}}

\newcolumntype{C}[1]{>{\centering\arraybackslash}m{#1}}
\newcolumntype{M}{>{\centering\arraybackslash$}l<{$}}

\newcommand{\eps}{\epsilon}
\def\trFive{{\rm tr}_5}

\newcommand{\MSbar}{\overline{\text{MS}}}

\newcommand{\ii}{\mathrm{i}}

\newcommand{\red}[1]{\textcolor{red!50!black}{#1}}

\begin{document}

\title{Double-Virtual NNLO QCD Corrections for Five-Parton Scattering:\protect\\ The Gluon Channel}
\preprint{PSI-PR-24-03, ZU-TH 75/23}


%
\author{Giuseppe~De~Laurentis}
\affiliation{Higgs Centre for Theoretical Physics, University of Edinburgh, Edinburgh, EH9 3FD, United Kingdom}
\affiliation{Paul Scherrer Institut, Forschungsstrasse 111, 5232 Villigen, Switzerland}
\author{Harald~Ita}
\affiliation{Paul Scherrer Institut, Forschungsstrasse 111, 5232 Villigen, Switzerland}
\affiliation{ICS, University of Zurich, Winterthurerstrasse 190, Zurich, Switzerland}
\author{Maximillian~Klinkert}
\affiliation{Physikalisches Institut, Albert-Ludwigs-Universit\"at Freiburg,
Hermann-Herder.~Str.~3, D-79104 Freiburg, Germany}
\author{Vasily~Sotnikov}
\affiliation{Physik-Institut, University of Zurich, Winterthurerstrasse 190, 8057 Zurich, Switzerland}

\begin{abstract}
We compute the two-loop helicity amplitudes for
the scattering of five gluons, including all
contributions beyond the leading-color approximation.
The analytic expressions are represented as linear combinations of transcendental functions with rational coefficients,
which we reconstruct from finite-field samples obtained with the numerical unitarity method.
Guided by the requirement of removing unphysical singularities, 
we find a remarkably compact generating set of rational coefficients, 
which we display entirely in the manuscript.
We implement our results in a public code,
which provides efficient and reliable numerical evaluations for phenomenological applications. 
\end{abstract}

\maketitle


\section{Introduction}\label{sec:introduction}

The long-term perspective of the Large Hadron Collider (LHC) at CERN
serves as a compelling reason to explore new ways to advance our
understanding of high-energy particle collisions beyond the level of
detail and precision currently attainable. 

Among the central
processes under study through hadron collisions is the production of multiple
jets.
Notably, the recent impressive measurements of the strong coupling
constant at high momentum
transfer \cite{Czakon:2021mjy,ATLAS:2023tgo,Alvarez:2023fhi} crucially
depend upon the cutting-edge next-to-next-to-leading order (NNLO) QCD
predictions for three-jet production \cite{Czakon:2021mjy} (see also
\cite{Chen:2022ktf}).  These predictions relied upon the leading-color
approximation for double-virtual corrections \cite{Abreu:2021oya},
contributing on average about $10\%$ \cite{Czakon:2021mjy}. 
This highlights the potential importance of including
subleading-color effects.
Moreover, the observation that the subleading-color
effects can be enhanced in certain differential
observables \cite{Chen:2022tpk} underscores the necessity to
study three-jet production in full color.
 
Looking towards the future, five-parton scattering at two loops
is also a crucial ingredient in advancing towards the
N\textsuperscript{3}LO precision frontier for di-jet production in
hadron collisions.  A related intriguing application 
lies in the explicit examination of the breakdown of
collinear factorization in QCD, which may occur at the third order in
perturbation theory when subleading color effects are taken into account
\cite{Catani:2011st,Forshaw:2012bi,Dixon:2019lnw}.

In this letter we focus on the five-gluon channel and
derive compact analytic expressions for all two-loop five-gluon
helicity amplitudes, including for the first time all contributions
beyond the leading-color approximation.  The remaining quark channels
will be presented in the followup publication \cite{DeLaurentis:2023izi}.

Five-point two-loop computations are notoriously difficult due to the
unforgiving admixture of algebraic, analytic, and combinatorial
complexity.  This challenge is particularly pronounced when
contributions from non-planar diagrams are included.  Nevertheless,
thanks to advancements in the understanding of the relevant Feynman
integrals
\cite{Papadopoulos:2015jft,Gehrmann:2018yef,Abreu:2018aqd,Chicherin:2018old,Chicherin:2020oor},
and analytic reconstruction techniques %
\cite{vonManteuffel:2014ixa,Peraro:2016wsq,Klappert:2020nbg,Magerya:2022hvj,Belitsky:2023qho,%
DeLaurentis:2022otd,Badger:2021imn,Abreu:2021asb,Abreu:2018zmy,Liu:2023cgs,DeLaurentis:2019bjh,%
DeLaurentis:2020qle,Campbell:2022qpq%
},
recent years have witnessed remarkable progress in computations of
two-loop five-point massless amplitudes. In fact, all massless two-loop amplitudes with any
combination of photons and partons in the final state are already known
analytically in full color
\cite{Agarwal:2021vdh,Badger:2023mgf,Badger:2021imn}, with the exception of the
five-parton process, which until now was known only in the
leading-color approximation \cite{Abreu:2021oya}.

In this work we build upon this remarkable progress and, in
particular, leverage the computational framework established in
refs.~\cite{Ita:2015tya,Abreu:2017xsl,Abreu:2017hqn,Abreu:2020xvt,Abreu:2023bdp}.
We further delve into a limiting
aspect of the analytic reconstruction methods in amplitude computations.
Conventional approaches,
in essence, employ generic rational ansätze
that involve numerous unphysical singularities and redundant parameters. 
In stark contrast amplitudes often assume concise representations in partial
fractioned form.
This leads to the intriguing question: can
analytic reconstruction directly yield the compact results?
While we delay a more thorough discussion of this question to future
work \cite{Wjjcompact}, we present evidence that a positive outcome can be
achieved by leveraging information about the functions' singularities and
residues.
In fact, considering two-loop five-gluon scattering in full color,
we obtain a representation 
that seamlessly fits into the appendix of this letter.
With the exception of special helicity
configurations \cite{Badger:2019djh,Dunbar:2019fcq,Kosower:2022bfv},
this level of simplicity has been notably elusive for
five-point two-loop scattering.

Finally, we provide a \texttt{C++} library for fast numerical
evaluation of the NNLO hard function that is ready for use in
cross-section computations.  Together with the upcoming results for the quark channels that we
will make available in the followup publication \cite{DeLaurentis:2023izi},
these results will provide crucial input
for NNLO cross-section computations.

\vspace{2em}
\textbf{Note added:} while this work was in preparation, we became aware of ref.~\cite{Agarwal:2023suw},
which reports partially overlapping results. 
We thank the authors for the numerical comparison of our results and for coordinating the publications.

\section{Notation and conventions}\label{sec:notation}

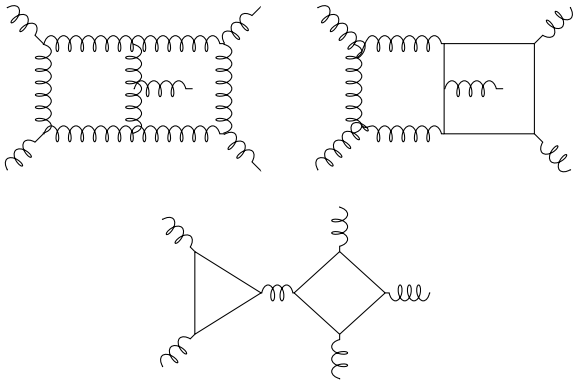
\begin{figure}[ht]
\centering
\begin{subfigure}{0.48\linewidth}
  \begin{tikzpicture}[scale=0.6]
    \node (g1) at (-1,1){};
    \node (g2) at (-1,-3){};
    \node (g3) at (5,1){};
    \node (g4) at (5,-3){};
    \node (g5) at (3.5,-1){};
    \coordinate (a1) at (0,0);
    \coordinate (a2) at (0,-2);
    \coordinate (b1) at (2,0);
    \coordinate (b3) at (2,-1);
    \coordinate (b2) at (2,-2);
    \coordinate (c1) at (4,0);
    \coordinate (c2) at (4,-2);
    \draw [gluon] (g1) -- (a1);
    \draw [gluon] (c1) -- (g3);
    \draw [gluon] (a1) -- (b1) -- (c1);
    \draw [gluon] (g2) -- (a2);
    \draw [gluon] (c2) -- (g4);
    \draw [gluon] (a2) -- (b2) -- (c2);
    \draw [gluon] (a1) -- (a2);
    \draw [gluon] (b1) -- (b2);
    \draw [gluon] (b3) -- (g5);
    \draw [gluon] (c1) -- (c2);
  \end{tikzpicture}
\end{subfigure}
~
\begin{subfigure}{0.48\linewidth}
  \begin{tikzpicture}[scale=0.6]
    \node (g1) at (-1,1){};
    \node (g2) at (-1,-3){};
    \node (g3) at (5,1){};
    \node (g4) at (5,-3){};
    \node (g5) at (3.5,-1){};
    \coordinate (a1) at (0,0);
    \coordinate (a2) at (0,-2);
    \coordinate (b1) at (2,0);
    \coordinate (b3) at (2,-1);
    \coordinate (b2) at (2,-2);
    \coordinate (c1) at (4,0);
    \coordinate (c2) at (4,-2);
    \draw [gluon] (g1) -- (a1) -- (b1);
    \draw [gluon] (a1) -- (a2); 
    \draw [quark] (b1) -- (c1) -- (c2)  -- (b2) -- (b1);
    \draw [gluon] (g3) -- (c1);
    \draw [gluon] (g2) -- (a2) -- (b2);
    \draw [gluon] (g4) -- (c2);
    \draw [gluon] (b3) -- (g5);
  \end{tikzpicture}
\end{subfigure}
\begin{subfigure}{0.48\linewidth}
  \begin{tikzpicture}[scale=0.55]
    \node (g1) at (-1,1){};
    \node (g2) at (-1,-3){};
    \node (g3) at (3.5,1.3){};
    \node (g4) at (3.5,-3.3){};
    \node (g5) at (5.9,-1){};
    \coordinate (a1) at (0,0);
    \coordinate (a2) at (0,-2);
    \coordinate (b1) at (1.6,-1);
    \coordinate (b2) at (2.4,-1);
    \coordinate (c1) at (3.5,0);
    \coordinate (c2) at (3.5,-2);
    \coordinate (d1) at (4.6,-1);
    \draw [gluont] (b1) -- (b2);
    \draw [gluon] (g1) -- (a1);
    \draw [gluon] (g2) -- (a2);
    \draw [gluon] (g3) -- (c1);
    \draw [gluon] (g4) -- (c2);
    \draw [gluont] (g5) -- (d1);
    \draw [quark] (a1) -- (b1) -- (a2) -- (a1); 
    \draw [quark] (b2) -- (c1) -- (d1) -- (c2) -- (b2); 
  \end{tikzpicture}
\end{subfigure}
\caption{Representative Feynman diagrams for two-loop five-gluon amplitudes. Solid lines represent closed massless quark loops.}
\label{tab:diagram_5g}
\end{figure}

%
We consider the $\order{\alpha_s^2}$ corrections to the scattering of
five-gluons.
This requires the computation of two-loop five-gluon scattering
amplitudes, which we obtain omitting the contributions from the massive
top quark. Furthermore, we treat all quarks as massless states.
The contributing partonic process is
\begin{equation}
\begin{gathered}\label{eq:5g}
 g(-p_1^{-h_1})+g(-p_2^{-h_2})\,\to\kern20mm \\ \kern20mm g(p_3^{h_3})+g(p_4^{h_4})+g(p_5^{h_5})\,.
\end{gathered}
\end{equation}
Here $p_i$ and $h_i$ denote the momentum and the helicity of the
$i^{\text{th}}$ particle, respectively. 
Unless stated otherwise, throughout this paper, momenta
and helicity labels are understood in the all-outgoing convention.
Representative Feynman diagrams for the two-loop contributions are shown in \cref{tab:diagram_5g}.

\subsection{Kinematics}

The process involves five massless particles. The underlying
scattering kinematic can therefore be specified by five Mandelstam invariants $\{s_{12},s_{23},s_{34},s_{45},s_{15}\}$,
as well as the parity-odd contraction of four momenta $\trFive = \trace(\gamma^5\slashed{p}_1\slashed{p}_2\slashed{p}_3\slashed{p}_4)$.

To represent the dependence of scattering amplitudes on the particles' helicities we use
two-component spinors, $\lambda_{i}^{\alpha}$ and $\tilde\lambda_{i}^{\dot\alpha}$, with $i \in \{1, \dots, 5\}$. 
We define the invariant contractions of spinors as
\begin{equation}
  \langle i j \rangle = \lambda^\alpha_i \lambda_{j, \alpha} \quad \text{and} \quad [ij] = \tilde\lambda_{i,\dot\alpha}\tilde\lambda_j^{\dot\alpha}\, ,
\end{equation}
which are related to the Mandelstam invariants through $s_{ij}=\langle ij\rangle [ji]$ (see e.g. \cite{Maitre:2007jq} for matching conventions). 
We will also use longer spinor
contractions, in particular
\begin{equation}
  \langle i | j \pm k | i ] = \langle i j \rangle [ji] \pm \langle ik
    \rangle [ki] \, .
\end{equation}
Finally, we can express $\trFive$ as a polynomial in spinor
brackets as\footnote{We note that $\trFive$ in ref.~\cite{Abreu:2021oya} differs by a minus sign compared to this definition.}
\begin{equation}
  \trFive = [12]\langle23\rangle[34]\langle41\rangle-\langle12\rangle[23]\langle34\rangle[41] \, .
\end{equation}

A little-group transformation of the $i^{th}$ leg with helicity $h_i$
reads $(\lambda_i,\tilde\lambda_i)\rightarrow
(z_i \lambda_i,\tilde\lambda_i/z_i)$. Under this transformation, the
helicity amplitudes transform as $A \rightarrow z_i^{-2h_i} A$. We
refer to the exponent of the $z_i$ as the little-group weight.

\subsection{Color space}\label{sec:amplitudes}
The external gluons are in the adjoint representation of $SU(N_c)$ and carry indices $\vec{a} = \{a_1,\ldots{},a_5\}$ which run over $N_c^2-1$ values. 
We explicitly represent the five-gluon amplitudes in the color space
through the trace basis as \cite{Bern:1990ux}
\begin{multline}\label{eq:partial_5g}
  \mathcal{A}_{\vec{a}}= \\
 \sum_{\sigma \in \mathcal{S}_5/\mathcal{Z}_5} \sigma\Big(\tr(1,2,3,4,5) \; A_{1}(1,2,3,4,5)\Big) \; + \\
 \sum_{\sigma\in \frac{\mathcal{S}_5}{\mathcal{Z}_2 \times \mathcal{Z}_3}} \sigma\Big(\tr(1,2) \tr(3,4,5) \; A_{2}(1,2;3,4,5)\Big),
\end{multline}
where $\tr(i_1,\ldots{},i_n) = \tr(T^{a_{i_1}}\cdots{}T^{a_{i_n}})$,
and $T^{a_{i}}$ are the hermitian and traceless generators of fundamental representation of $SU(N_c)$.
The permutation $\sigma = \{i_1,\ldots{},i_5\}$ acts on all external-particle labels as $\sigma(i) = \sigma_i$.
The sums run over all permutations that do not leave the respective traces invariant.
Thus, the first sum runs over 24 elements, while the second one runs over 20 elements.

The generators $T^a$ are normalized as,
\begin{equation}
{\tr}(T^a T^b) = \delta^{ab}\,,
\end{equation}
and fulfill the commutator relations,
\begin{align}
  \left[T^a,T^b\right] & =i f_{abc} T^c \, , \\
  i f_{abc} & =\tr(T^a T^b T^c) - \tr(T^b T^a T^c) \,.
\end{align}

The amplitudes $A_i$ admit an expansion in terms of the bare QCD
coupling constant $\alpha_s^0 = (g_s^0)^2/(4\pi)$,
\begin{equation}\label{eq:as-series}
  A_i = (g_s^0)^3 \left(\sum_{L=0}^2 \left(\frac{\alpha_s^0}{2 \pi}\right)^L A_i^{(L)} ~+~ \order{(\alpha_s^{0})^3}\right) 
\end{equation}
with $L$ denoting the number of loops. 
In this work we consider an arbitrary number of light quark flavors in the loops that is denoted by $N_f$.
The five-gluon amplitudes can be further expanded in powers of $N_c$ and $N_f$ through two loops as follows,
\begin{subequations}\label{eq:Nc-Nf-series}
\begin{align}
  A_1^{(0)} &= A^{(0),(0,0)} \,, \qquad A_2^{(0)} = 0\,,\\[1em]
  A_1^{(1)} &= N_c~A^{(1),(1,0)} ~+~ N_f~A^{(1),(0,1)}\,, \\ 
  A_2^{(1)} &= A^{(1),(0,0)}\,, \\[1em]
  A_1^{(2)} &= N_c^2~A^{(2),(2,0)} ~+~ \red{A^{(2),(0,0)}}\, \\ \nonumber
            & +~ N_f N_c ~A^{(2),(1,1)} ~+~ \frac{N_f}{N_c} ~ \red{A^{(2),(-1,1)}}\, \\ \nonumber
            & +~ N_f^2 ~A^{(2),(0,2)}\,, 
  \\
  A_2^{(2)} &= N_c~ \red{A^{(2),(1,0)}} \\ \nonumber
            & +~ N_f~ \red{A^{(2),(0,1)}} ~+~ \frac{N_f^2}{N_c} ~ \red{A^{(2),(-1,2)}}\,.
\end{align}
\end{subequations}
The coefficients $A^{(L),(n_c,n_f)}$, which we call \emph{partial amplitudes}, are uniquely identified by the three integers $L,n_c,n_f$.
Therefore, to avoid clutter we omit the subscripts on the right hand side of \cref{eq:Nc-Nf-series}.
In the limit of large number of colors with $N_f/N_c$ fixed only the partial amplitudes with $L=n_c+n_f$ contribute.
These leading-color partial amplitudes receive contributions only from planar diagrams \cite{tHooft:1973alw} and have been calculated in \cite{Bern:1993mq,Badger:2015lda,Badger:2018enw,Abreu:2018zmy,Abreu:2019odu},
while $A^{(2),(1,0)}$ and $A^{(2),(0,0)}$ are known in the special all-plus helicity configuration only \cite{Badger:2019djh,Dunbar:2019fcq,Kosower:2022bfv}.
The amplitudes $\red{A^{(2),(0,0)}}$, $\red{A^{(2),(-1,1)}}$, $\red{A^{(2),(1,0)}}$,  $\red{A^{(2),(0,1)}}$, $\red{A^{(2),(-1,2)}}$ receive contributions from non-planar diagrams and are the new result of this work.
We note that the amplitudes $A^{(2),(-1,2)}$ vanish for any helicity assignments.
For convenience we also recalculate all previously known amplitudes in \cref{eq:Nc-Nf-series}.

\subsection{Renormalization}

We regularize ultraviolet (UV) and infrared (IR) divergences of loop amplitudes in the 't~Hooft--Veltman scheme of dimensional regularization, setting the
space-time dimensions to $D=4-2\epsilon$.
The UV divergences are removed by renormalization of the bare QCD coupling in the 
$\MSbar$ scheme. To achieve this we perform the following replacement in \cref{eq:as-series},
\begin{align}\label{eq:renorm}
  &\alpha_0\mu_0^{2\epsilon}S_{\epsilon}
  =\alpha_s\mu^{2\epsilon} \times \\
  &\left(
  1-\frac{\beta_0}{2\epsilon}\frac{\alpha_s}{2\pi}
  +\left(\frac{\beta_0^2}{4 \epsilon^2}-\frac{\beta_1}{ 8 \epsilon}\right) \left(\frac{\alpha_s}{2\pi}\right)^2+\mathcal{O}\left(\alpha_s^3\right)\right)\,, \nonumber
\end{align}
where $S_\epsilon=(4\pi)^{\eps}e^{-\eps\gamma_E}$, with
$\gamma_E=-\Gamma'(1)$ the Euler-Mascheroni constant, 
and $\mu_0,\mu$ are regularization and renormalization scale respectively.
The QCD $\beta$-function coefficients are
\begin{subequations}
  \begin{align}
    \beta_0&=\frac{11}{3} N_c -  \frac{2}{3} N_f \;\, , \\
    \beta_1&=\frac{34}{3} N_c^2 - \frac{13}{3} N_c N_f + \frac{N_f}{N_c} \,.
  \end{align}
\end{subequations}
The renormalized amplitudes are expanded through the renormalized coupling as in \cref{eq:as-series}.

The remaining infrared divergences can be extracted through the universal factorization 
\cite{Catani:1998bh,Sterman:2002qn,Becher:2009cu,Gardi:2009qi}:
\begin{equation}\label{eqn:remainder}
  {\cal R} ={\bf Z}(\epsilon, \mu) {\cal A} ~+~ \order{\epsilon} \,,
\end{equation}
where the \emph{finite remainder} ${\cal R}$ is obtained through the application of
the color-space operator ${\bf Z}$. The latter is obtained \cite{Becher:2009cu} from the path-ordered evolution 
\begin{align}
{\bf Z}^{-1} (\epsilon, \mu)={\bf P}\, {\rm exp}\left[\int^\infty_{\mu}\frac{d\mu^\prime}{\mu^\prime} {\bf \Gamma}(\mu^\prime)\right],
\end{align}
of the anomalous dimension matrix
\begin{align} \nonumber
&\kern-5mm{\bf\Gamma}(\mu)= -\sum_{(i,j)} {\bf T}_i\cdot {\bf T}_j \times \\
&\kern5mm\frac{\gamma_{\rm cusp}}{2} \; {\rm ln}\left(-\frac{s_{ij}}{\mu^2} - \ii 0\right) ~+~ 5 \gamma^g ,
\end{align}
where the sum runs over all pairs of external gluons,
and the color operators $\mathbf{T}_i$ act on the color representation of the
i$^{th}$ parton. 
For adjoint indices the action is given by $({\bf T}_i^a)_{bc}=-i f^{abc}$.
The anomalous dimensions $\gamma_{\mathrm{cusp}}$ and $\gamma^g$ can be found in \cite[Appendix A]{Becher:2009qa}\footnote{
  We rescale them by a factor of 2 per loop to match our expansion in $\alpha_s/2\pi$ in \cref{eq:as-series}.
}.

After UV and IR renormalization of amplitudes through \cref{eq:renorm,eqn:remainder}
we recover expansions of \cref{eq:as-series,eq:partial_5g,eq:Nc-Nf-series} for the finite remainders $\mathcal{R}$,
and therefore obtain \emph{partial finite remainders}
\begin{align}
  R_{\vec{h}}^{(L),(n_c,n_f)}(i_1,\ldots{},i_5),
\end{align}
which are the elementary building blocks that we focus on in this work.
It is worth noting that the finite remainders contain the complete physical information about the underlying scattering process. 
In particular, any observable can be calculated through finite remainders (see e.g.~\cite{Weinzierl:2011uz}),
which allows one to cancel much of the disruptions caused by the use of dimensional regularization.
Finite remainders in a different IR renormalization scheme (with a different operator $\mathbf{Z}$) can be obtained by an additional \emph{finite} renormalization
after \cref{eqn:remainder}. We elaborate on this in the forthcoming publication \cite{DeLaurentis:2023izi}.

\subsection{Generating set of finite remainders}
\label{subsec:generating-partial-helicity-remainders}

To calculate arbitrary observable quantities we must know all partial remainders from \cref{eq:Nc-Nf-series} 
in all permutations in \cref{eq:partial_5g}, with $2^5$ helicity assignments for each of them.
Fortunately, the combinatorial complexity can be sidestepped by mapping each partial helicity remainders
onto a small generating set.

First, it is well known that in the Yang-Mills theory partial
amplitudes (in the trace basis) satisfy additional identities (see
e.g.\ \cite{Edison:2011ta} for a systematic study,
and \cite{Dunbar:2023ayw} for a recent review).  We verified by a
direct computation that the relations between two-loop partial
amplitudes discussed in ref.~\cite{Edison:2011ta} hold both at the
level of amplitudes and finite remainders.  They allow us to express
$R^{(2),(0,0)}$ through sums over permutations of $R^{(2),(2,0)}$ and
$R^{(2),(1,0)}$ for each helicity assignment.  Interestingly, we find
no linear relation among permutations of $R^{(2),(1,1)}$,
$R^{(2),(0,1)}$ and $R^{(2),(-1,1)}$.

Next, we work out the generating set for each partial remainder constructively by starting with the set of all helicities and permutations,
and partitioning it into the orbits under the action of the group $\mathcal{P}\otimes\mathcal{C}\otimes\Sigma_i$,
where $\mathcal{P}$ and $\mathcal{C}$ are parity- and charge-conjugation respectively, and $\Sigma_i$ is the symmetry group of the corresponding 
color structure in \cref{eq:partial_5g}. 
We can then pick one representative from each orbit, and obtain the complete set of remainders by relabeling momenta and symmetries.
Let us note, that for the reasons that will become clear in \cref{sec:reconstruction}, we do not choose the identity permutation $\{1,2,3,4,5\}$ for all our representatives.
Instead we prioritize to have uniform spinor weight for each of the three helicity assignments $+++++$ (all-plus), $++++-$ (single-minus), and $+++--$ (MHV).

For the all-plus helicity configuration we have the generating remainders
\begin{subequations} \label{eqn:RallPlus}
  \begin{align}
    & R_{1}(1^+,2^+,3^+,4^+,5^+) \, , \\
    & R_2(1^+,2^+;3^+,4^+,5^+). \,
  \end{align}
\end{subequations}
For the single-minus helicity configuration we have 
\begin{subequations}\label{eqn:RsingleMinus}
\begin{align}
&R_1(1^+,2^+,3^+,4^+,5^-) \,,\\
&R_2(1^+,2^+;3^+,4^+,5^-) \,,\\
&R_2(1^+,5^-;4^+,3^+,2^+) \,.
\end{align}
\end{subequations}
Finally, for the MHV configurations we have five generating remainders,
\begin{subequations}\label{eqn:Rmhv}
  \begin{align}
  &R_1(1^+,2^+,3^+,4^-,5^-) \,,\\
  &R_1(1^+,2^+,4^-,3^+,5^-) \,,\\
  &R_2(1^+,2^+;3^+,4^-,5^-) \,,\\
  &R_2(1^+,5^-;2^+,3^+,4^-) \,,\\
  &R_2(5^-,4^-;3^+,2^+,1^+) \,.
  \end{align}
\end{subequations}
Here we suppress the labels $L,n_c,n_f$, and, for better readability, we show the helicity
labels as superscripts of the momentum labels.

\subsection{NNLO hard function}

To calculate the double-virtual contribution to NNLO QCD partonic cross sections 
one must square the helicity finite remainder $\mathcal{R}_{\vec{h},\vec{a}}$ in \cref{eq:partial_5g}
and perform summation over color and helicity indices.

At leading order in $\alpha_s$ we define the function
\begin{equation}\label{eq:born}
  \mathcal{B} =  \sum_{\vec h, \vec a} \abs{\mathcal{A}^{(0)}_{\vec{h},\vec{a}}}^2 = N_c^3 (N_c^2 -1) ~ B \,.
\end{equation}
We then define the \emph{hard function} $\mathcal{H}$ as
\begin{equation}\label{eq:hard-function}
  \mathcal{H}= \frac{1}{\mathcal{B}} \sum_{\vec h, \vec a} \abs{{\cal R}_{\vec h,\vec a}}^2 \,,
\end{equation}
which can be expanded perturbatively up to $\order{\alpha_s^2}$ as in \cref{eq:as-series}. 
Similar to \cref{eq:Nc-Nf-series}, we can expand $\mathcal{H}$ in powers of $N_c$ and $N_f$. 
Through two loops we get
\begin{subequations}
  \label{eq:hard-function-partial}
  \begin{align}
    \mathcal{H}^{(0)} &= 1 \,,  \\
    \mathcal{H}^{(1)} &= N_c ~ H^{(1),(1,0)} + \frac{1}{N_c}~H^{(1),(-1,0)} \nonumber \\
                      &+   N_f ~ H^{(1),(0,1)}  +   \frac{N_f}{N_c^2}~H^{(1),(-2,1)} \,, \\
    \mathcal{H}^{(2)} &= N_c^2 ~ H^{(2),(2,0)} + ~H^{(2),(0,0)} \,, \nonumber \\
                      & +   N_f  \sum_{n_c\in \{1,-1,-3\}} N_c^{n_c} ~ H^{(2),(n_c,1)} \,, \nonumber\\ 
                      & +   N_f^2  \sum_{n_c\in \{0,-2,-4\}} N_c^{n_c} ~ H^{(2),(n_c,2)} \,. 
  \end{align}
\end{subequations}
Here again only the functions $H^{(L),(n_c,n_f)}$ with $L=n_c+n_f$ contribute in the leading-color approximation.

To perform the summations in \cref{eq:hard-function} we use the maps inverse to the ones that were used in the previous section to construct the generating set of finite remainders. 
It is worth noting that the polynomial expansion in $N_c$ and $N_f$ given in \cref{eq:hard-function-partial} holds only when identities between partial remainders are correctly taken into account.

\section{Analytic reconstruction}
\label{sec:reconstruction}

We now discuss the computation of the finite remainders (\ref{eqn:remainder}).
They can be represented as a linear
combination of transcendental integral functions $h_i$ and rational coefficient
functions $r_i$,
\begin{equation}
R = \sum_i r_i h_i\, .
\end{equation}
For the integral functions $h_i$, we employ the set of non-planar
pentagon functions from ref.~\cite{Chicherin:2020oor}.  The
rational coefficient functions are the central result of this paper
and we obtain them via analytic reconstruction, i.e.~we start with an
ansatz for the rational coefficient functions $r_i$ and determine its
parameters from exact numerical evaluations of the remainder over prime fields.  
The reconstruction cost is dominated by the time of sampling the remainders. This
motivates us to search for a strategy to constrain the ansatz using physics
arguments and reduce its free parameters.  We will now discuss the details of this computation.

\subsection{Numerical sampling}
\label{sec:numEval}

For numerical amplitude evaluations in a finite field we use the program \textsc{Caravel} \cite{Abreu:2020xvt},
which implements the multi-loop numerical
unitarity method \cite{Ita:2015tya,Abreu:2017xsl,Abreu:2017hqn}.
In this approach amplitudes are reduced to a set of master integrals 
by matching numerical evaluations of generalized unitarity cuts to a parametrization of 
the loop integrands. For the five-gluon process we use the recently obtained 
parametrization \cite{Abreu:2023bdp}, which we extend in loop-momentum degree to 
match the corresponding dependence of the gluon cut diagrams. Furthermore,
we extended the set of planar unitarity cuts to non-planar diagrams which are required for 
subleading-color partial amplitudes.
We generated the cut diagrams with \texttt{qgraf}
\cite{Nogueira:1991ex}
and arranged them into a hierarchy of cuts with a private code. 
We matched the cuts evaluated through
color-ordered tree amplitudes to the amplitude definitions 
in section \ref{sec:amplitudes}, by employing the 
unitarity based color decomposition \cite{Ochirov:2016ewn,Ochirov:2019mtf}.
We extracted the $\epsilon$-dependence of cuts that
originates from the state sums in loops through the dimensional reduction method \cite{Badger:2017jhb,Abreu:2018jgq}.
 
With these extensions \textsc{Caravel} now computes the integral coefficients 
$r_i$ of five-gluon partial amplitudes up to two loops, given a kinematic point and a choice of polarization 
labels for the external gluons. 

Analytic expressions for the coefficient functions $r_i$ can then be
reconstructed using multivariate functional reconstruction techniques
\cite{vonManteuffel:2014ixa,Peraro:2016wsq} 
(see also recent
refs.~\cite{Klappert:2020nbg,Magerya:2022hvj,Belitsky:2023qho,Liu:2023cgs}) based
on Newton and Thiele's interpolation algorithms. 
However, our approach is in fact closer to the ansatz-based
approach of refs.~\cite{DeLaurentis:2019bjh,DeLaurentis:2022otd}.
This approach constructs an ansatz for the rational integral coefficients $r_i$,
which is constrained by information from the neighborhood of their singularities.
Nevertheless, here we differ even from this approach in that we use
information about the residues to build linear transformations of
rational functions $r_i$ to bases of functions $\tilde r_i$ with a
simplified pole structure and fewer ansatz parameters. This basis
change is determined numerically and simplifies the subsequent ansatz
construction and parameter determination.

We will require two types of numerical evaluations for the 
remainder functions:
\begin{enumerate}
\item Random phase-space points: these are $N$ randomly generated phase-space
points which we label by the superscript $n$. We represent these points in terms of 
sets of spinor variables,
\begin{align}\label{eqn:randomPS}
\{ \{\lambda_1^{n}, ... , \lambda_5^{n},\tilde \lambda_1^{n}, ... , \tilde \lambda_5^{n} \} \}_{n=1,N} \,.
\end{align}
They are subject to momentum conservation $\sum_i \lambda_i^n\tilde\lambda^n_i=0$.

\item A family of (anti-)holomorphic slices:
these are $\tilde N<N$ holomorphic slices 
\cite{PageSAGEXLectures,Abreu:2021asb,Abreu:2023bdp}
associated to a subset of
the random phase-space points (\ref{eqn:randomPS}),
\begin{equation}
\begin{aligned}\label{eqn:famHolomorphicSlice}
&\lambda_i^{\tilde n}(t)=\lambda_i^{\tilde n}+t c_i^{\tilde n} \eta^{\tilde n}\,,\quad 
\tilde\lambda_i^{\tilde n}(t) = \tilde\lambda_i^{\tilde n}\,,\\ 
&\qquad \sum_{i=1}^5 c_i^{\tilde n} \tilde\lambda_i^{\tilde n}=0\,.
\end{aligned}
\end{equation}
Here the reference spinor $\eta^{\tilde n}$ is chosen randomly. The label $\tilde n$ runs 
over $\tilde N$ values.
Similarly we will use anti-holomorphic slices which are obtained from
(\ref{eqn:famHolomorphicSlice}) by swapping $\lambda_i^{\tilde n}\leftrightarrow
\tilde\lambda_i^{\tilde n}$ and $\eta^{\tilde n} \rightarrow \tilde\eta^{\tilde n}$.
\end{enumerate} 

\subsection{Coefficient-function basis}
\label{sec:CoeffBasis}

We now identify a basis of coefficient functions $r_i$ based on a measure for
the functions' complexity.  We start from the general form of the rational
coefficient functions in spinor variables,
\begin{equation} \label{eqn:LCD}
r_i = \frac{{\cal N}_i(\lambda,\tilde \lambda)}{\prod_{j}
{\cal D}_j^{q_{ij}}(\lambda,\tilde \lambda)}\,.  
\end{equation} 
The denominator factors ${\cal D}_j$ are  given by the symbol alphabet with 
integer exponents $q_{ij}$ \cite{Abreu:2018zmy}. 
Computing $r_i$ amounts to determining the $q_{ij}$ and all parameters in the
numerator polynomial ${\cal N}_i$. While it is straightforward to determine 
$q_{ij}$ (see below) it is non-trivial to determine ${\cal N}_i$ 
because of the typically large polynomial degree. 
Consequently, a useful measure of complexity is the mass dimension of ${\cal
N}_i$, which is linked to the number of parameters the polynomial depends on. 

Simple dimensional analysis allows us to determine the mass dimension and spinor weights of the
numerators ${\cal N}_i$, from those of the polynomials ${\cal D}_j$,
the data $q_{ij}$ and the overall mass dimension and spinor weights of
the helicity remainder. Consequently, we only need to determine the exponents $q_{ij}$,
which we compute following
the univariate-slice reconstruction \cite{Abreu:2018zmy} in
spinor-helicity variables \cite{PageSAGEXLectures} (see also
refs.~\cite{Abreu:2021asb,Abreu:2023bdp}).  Here the remainder
functions are reconstructed on a single holomorphic and a single
anti-holomorphic slice (\ref{eqn:famHolomorphicSlice}), and
subsequently the denominators are matched to products of the letter
polynomials ${\cal D}_j(t)$.  This uniquely fixes the exponents
$q_{ij}$ for each function $r_i$.

For the five-gluon finite remainders we observe that the set of denominator factors contains the $35$ elements,
\begin{align}\label{eq:codimension-one-poles}
  {\cal D} =& 
  \big\{\langle ij \rangle, 
	\, [ij], \,
	\langle i|j+k|l] ,... \big\}
\end{align}
where the set runs over all independent permutations of the spinor
strings/brackets. We also note that no coefficient in the finite remainder has a $\text{tr}_5$ singularity. 
From the $q_{ij}$ we can immediately deduce the mass dimension of the numerators ${\cal N}_i$. 

Our next goal is to determine a minimal function basis for the set of coefficient
functions $r_i$ in order to reduce the number of rational functions $r_i$ that
must to be constructed \cite{Abreu:2018jgq}.  To this end, we exploit the
freedom in our choice of basis functions and select a basis containing numerators with the lowest
mass dimensions. 
We express the linear dependent $r_i$ through the basis as
\begin{align} r_i =
\sum_{j\in{\rm basis}} r_j  M_{ji} \,, 
\end{align} 
where $M_{ij}$ is a constant rectangular matrix.
At this point we do not have analytic expressions for the $r_i$
available and we require a numerical method to identify linear
dependence of functions.  This is achieved using a set of random
evaluations (\ref{eqn:randomPS}) \cite{Abreu:2018rcw,Abreu:2018jgq},
which allows to associate vectors $\vec r_i$ of function values to
each $r_i$. A function basis is then identified by identifying a basis
in the vector space of function values $\vec r_i =\sum_{j\in {\rm
basis}} \vec{r}_j M_{ji}$ by linear algebra.  After this step, we
arrive at the form of the remainder,
\begin{equation} R = \sum_{j\in {\rm basis}, i} r_j M_{ji} h_i \, .
\end{equation}
Obtaining the analytic form of the remainder now amounts to 
computing the set $\{r_i\}_{i\in {\rm basis}}$.

\subsection{Basis change}
\label{sec:basisChange}

So far we have selected a convenient basis of functions $\{r_i\}_{i\in {\rm
basis}}$. Next, we will exploit universal properties of the functions' poles,
namely correlations between residues, to construct linear transformations to a
simpler basis of functions, which we will denote by $\tilde r_i$. (`Simpler' again
refers to the $\tilde r_i$ functions having numerators with lower mass
dimension than the ones of the $r_i$.) A basis change of this type was used
already some time ago in ref.~\cite{Abreu:2018zmy} and a detailed discussion 
of the  algorithm will be presented in the forthcoming paper \cite{Wjjcompact}.
Here we summarize
the main steps.  The reason that such linear combinations exist, stems from the
fact that many poles (at zeros of the ${\cal D}_i$) are spurious and cancel in
the remainders.  (Examples of spurious-pole denominators are spinor strings $\langle
i|j+k|l]$, as well as higher-order spinor products $[ij]$ and $\langle ij
\rangle$ which do not contribute to factorization poles.) Such cancellations
require that the residues of distinct $r_i$ are linearly related;
only in this way they may cancel
when multiplied with (degenerate) transcendental functions $h_i$ evaluated on
the respective singular surfaces.

We now discuss the simplifying basis change from the functions $r_i$ to $\tilde
r_i$ for a given remainder.
We start by defining our requirement for a constant linear basis change 
$O_{ij}$,
\begin{align}
\tilde r_i = \sum_{j\in {\rm basis}} O_{ij} r_j\,.
\end{align}
Our objective is to ensure that the mass dimension of all new numerators $\tilde{\cal N}_{i}$,
\begin{align}\label{eqn:LCDimproved}
\tilde r_i = \frac{\tilde {\cal N}_i(\lambda,\tilde \lambda)}{\prod_{j}
{\cal D}_j^{\tilde q_{ij}}(\lambda,\tilde \lambda)}\,,
\end{align}
is lower than those in the original least common denominator (LCD) 
form (\ref{eqn:LCD}),
\begin{align}
{\rm dim}( \tilde {\cal N}_j) < {\rm dim}( {\cal N}_i ) \quad i\in {\rm basis}
\end{align}
where ${\rm dim}({\cal N}_i)$ denotes the mass dimension of 
numerator ${\cal N}_i$.

The property we use is that the mass dimension is linked to the exponents $\tilde q_{ij}$. 
We are thus lead to consider the residues of the coefficient functions on the zeros of 
the denominator factors ${\cal D}_j$.
To this end we use a family of holomorphic univariate slices (\ref{eqn:famHolomorphicSlice})
and obtain a univariate representation of the coefficient functions $\tilde r_i(t)$\,. 
Near a single zero $t_k$ of one of the denominator factors 
\begin{align}
{\cal D}_k(t_k)=0, 
\end{align}
we obtain a formal Laurent series
\begin{align}\label{eqn:residue}
r_i(t)=\sum_{m=1}^{q_{ik}} \frac{e^k_{im}}{(t-t_k)^m} + {\cal O}\left((t-t_k)^0\right)\, ,  
\end{align}
with $e^k_{im}$
being functions of the external kinematics.
The $e^k_{im}$ will be referred to as codimension-one residues. 
To write the above equation (\ref{eqn:residue}) in a more uniform way, 
we introduce the maximal pole degree of a given factor ${\cal D}_k$,
\begin{align}
q_k = {\rm max}_{i\in {\rm basis}} [ q_{ik}] \,.
\end{align}
At the same time we introduce vanishing residues,
\begin{align}
e^k_{im} = 0 \quad \mbox{for}\quad  q_{ik }<m \le  q_k \,,
\end{align}
and rewrite (\ref{eqn:residue}) as,
\begin{align}\label{eqn:residuePrime}
r_i(t)=\sum_{m=1}^{q_{k}} \frac{e^k_{im}}{(t-t_k)^m} + {\cal O}\left((t-t_k)^0\right)\,.
\end{align}
In this way the summation is independent of the residue function and the degree of 
the pole is encoded in the vanishing of residues.
 
Equipped with this notation, we now return to the discussion of the constant
transformation matrix $O_{ij}$. We will now constrain the matrix, such that some
of the leading residues vanish for the new basis $\tilde r_i$. This is equivalent 
to lowering the respective pole degrees $\tilde q_{ij}$ (and consequently the numerators' mass dimensions),
which is what we aimed to achieve in the first place. 
The Laurent expansion of the functions $\tilde r_i$ depends linearly on 
the data of the Laurent expansion of $r_i$,
\begin{align}
\tilde r_i(t)&=\sum_{m=1}^{q_{k}} \frac{\tilde e^k_{im}}{(t-t_k)^m} + {\cal O}\left((t-t_k)^0\right)\,,\\
\tilde e^k_{im} &= \sum_{j\in {\rm basis}} O_{ij} e^k_{jm}  \,.
\end{align}
Consequently, we can choose the basis change $O_{ij}$ in such a way, 
that the leading residue vanishes, effectively reducing the power of the leading pole, 
\begin{align}
\sum_{j\in {\rm basis}} O_{ij} e^k_{j  q_{k}} = 0 \,.
\end{align}
We find that the requirement for a good basis change is that the rows
of the matrix $O_{ij}$ are in the kernel of the leading residue
functions.  Independently, we also have to ensure that the basis
change is invertible.  Given this property, the pole degrees of the
$\tilde r_i$ are reduced.

Finally, we adjust the criterion for the fact that we do not have analytic
expressions for the residue functions available. We follow the strategy of
ref.~\cite{Abreu:2018rcw,Abreu:2018jgq} (see also section \ref{sec:CoeffBasis}).
We repeat the above construction 
for $\tilde D$ holomorphic slices (\ref{eqn:famHolomorphicSlice}) 
(labeled by $\tilde n$) and obtain a representation of the residue functions 
$e^k_{im}$ as vectors of their function values 
\begin{align}
\vec e^{\,k}_{im}=\{ e^k_{im}(t^1_k) ,..., e^k_{im}(t^{\tilde D}_k)\}. 
\end{align}
The linear constraint for the basis change $O_{ij}$ is upgraded to,
\begin{align} \label{eqn:vanishRes}
\sum_{j\in {\rm basis}} O_{ij} \vec e^{\,k}_{j  q_{k}} = 0 \,,
\end{align}
and relies solely on numerical input.  The required number $\tilde D$
of slices is given by the dimension of the largest vector space of
codimension-one residues $e^{\,k}_{j q_{k}}$. Each row of $O_{ij}$
lives in the nullspace of $\vec e^{\,k}_{j q_{k}}$.

So far we were concerned with removing the leading singularity
associated to the pole labeled by $k$. In reality, we impose the
linear constraint (\ref{eqn:vanishRes}) for multiple residues
simultaneously. This includes originally subleading residues, if they
transform into leading residues following the cancellation of the
previous leading ones. In fact, for the most complicated function, $k$
and $m$ run over up to $40$ residues and orders. Furthermore, we find
it convenient to derive the basis change by considering one function
$\tilde r_i$ at a time, which amounts to constructing $O_{ij}$ row by
row. We start from the simplest functions, progressing towards the
most complex ones.

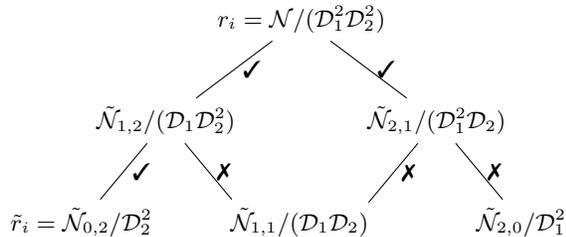
\begin{figure}
\hspace{-3mm}
\begin{tikzpicture}
[
        scale=0.9,
	level 1/.style = {black, sibling distance = 4cm},
	level 2/.style = {black, sibling distance = 2.5cm},
	level 3/.style = {black, level distance = 1cm},
	edge from parent path = {(\tikzparentnode\tikzparentanchor) -- (\tikzchildnode\tikzchildanchor)}
]

\node {$r_i = \mathcal{N}/(\mathcal{D}_1^2\mathcal{D}_2^2)$}
      child {
           node {$\tilde{\mathcal{N}}_{1,2}/(\mathcal{D}_1\mathcal{D}_2^2)$}
            child {node {$\tilde{r}_i = \tilde{\mathcal{N}}_{0,2}/\mathcal{D}_2^2$}
                  edge from parent node [right] {\cmark} }
            child {node(shared) {$\kern13mm\tilde{\mathcal{N}}_{1,1}/(\mathcal{D}_1\mathcal{D}_2)$}
                  edge from parent node [right] {\xmark} }
	edge from parent node [right] {\cmark}
        }
      child {
            node(2ndparent) {$\tilde{\mathcal{N}}_{2,1}/(\mathcal{D}_1^2\mathcal{D}_2)$}
	    child {node {$\phantom{\tilde{\mathcal{N}}}$}
                  edge from parent node [right] {\xmark} }  
	    child {node {$\tilde{\mathcal{N}}_{2,0}/\mathcal{D}_1^2$}
                  edge from parent node [right] {\xmark} }
	edge from parent node [right] {\cmark}
      };
\end{tikzpicture}
\caption{\label{FigSearchTree}Example of a simple search tree for intersecting 
null spaces with a global ($\tilde r_i$) and a 
local minimum ($\tilde{\cal N}_{2,1}/({\cal D}_1^2{\cal D}_2)$).}
\vspace{-2mm}
\end{figure}

In practice, to build each row, we perform an exhaustive breadth-first
search in the space of intersections of nullspaces of the $\vec
e^{\,k}_{im}$. That is, we build $\tilde r_i$ by attempting to remove
an increasingly higher number of poles, until no combination exists
that drops more. \Cref{FigSearchTree} shows a simplified example of
such a search tree, in the case where the function has only two
denominator factors. The tree is truncated at depth 2. A tick (\cmark)
or cross (\xmark) on the edge labels whether a linear combination of
the $r_i$'s exists that removes the given denominator factor. We label
the numerators by their respective denominator factor powers. We look
for a global minimum in the search tree. Although the $\mathcal{D}_i$
may differ in mass dimension, we find it most convenient to minimize
the sum of their exponents $\tilde{q}_{ij}$.

To ensure the rank stays maximal, after each row of $O_{ij}$ has been
computed, the following search for $\tilde r_{i+1}$ is done using
$\tilde r_{i'\leq i}$ and $r_{i'> i}$. A suitable pivoting strategy in
the computation of the nullspace intersections ensures $\tilde
r_{i+1}$ remains a pivot, and hence the rank does not drop. Otherwise,
the linear combination is discarded.  In certain cases, the
computation for the last rows of $O_{ij}$ was unnecessary, as these
functions were derived through symmetries (as discussed below) or
other partial remainders. The impact of the basis change is presented
in \cref{tab:ansatze-sizes} in terms of LCD ansatz sizes.

Using basis changes that decorrelate the functions poles we have achieved 
much simplified coefficient functions prior to performing a prohibitively expensive reconstruction.
The effect is somewhat similar to 
what was achieved by univariate partial fractioning by a suitable variable 
\cite{Badger:2021imn,Badger:2021nhg,Abreu:2021asb}.
However, we avoid introducing new spurious denominators, and in
addition the order of unphysical poles is systematically reduced. Both
effects are expected to lead to much improved numerical stability.
Moreover, we also note that, following the change in the set of basis
functions, 
further simplification may be achieved by a 
partial-fraction decomposition, either by the
semi-numerical slicing techniques just mentioned, or through the
purely numerical approach of refs.~\cite{DeLaurentis:2019bjh,
DeLaurentis:2022otd}.

\subsection{Analytic reconstruction and simplification}
\label{sec:reconstruction}

We are now in a position to reconstruct the coefficient functions. We
do this by constructing an ansatz matching the pole structure of the
final LCD form (\ref{eqn:LCDimproved}).  Given previous experience, we
further simplify the ansatz by removing terms with more than one
$\langle i | j+ k | i]$-factors  in the
denominator \cite{Abreu:2023bdp}. For example, we expect (and verify) $\mathcal{N}$
to be such that the following type of equality holds
\begin{equation}
\begin{gathered}
\kern-20mm \frac{{\cal N}}{\langle 4 | 1+ 5 | 4]^2 \langle 5 | 3 + 4 | 5]^2} = \\ 
\qquad \frac{{\cal N}_a}{\langle 4 | 1+ 5 | 4]^2} + \frac{{\cal N}_b}{\langle 5 | 3 + 4 | 5]^2}\,
\end{gathered}
\end{equation}
irrespective of what other denominator factors appear and of the degree of the spinor chains. This allows
us to make the ansatz in terms of the two lower-degree numerator
polynomials ${\cal N}_{a,b}$ instead of ${\cal N}$.  Constructing an
ansatz for the numerator polynomial can be non trivial.  We build
ansätze for the numerator polynomials in terms of independent
monomials of spinor brackets, which have both the right mass dimension
and little group weights. Mathematically, this amounts to the
enumeration of members of a polynomial quotient ring, subject to
irreducibility by a Gröbner basis and degree bounds \cite[section
2.2]{DeLaurentis:2022otd}. The ansatz construction relies on the
open-source programs \texttt{Singular} \cite{DGPS}, for the Gröbner
basis computation, and \texttt{OR-tools} \cite{ortools}, for the
linear programming.  Finally, we determine the numerator polynomials
by solving linear systems for a sufficient number of random
evaluations (\ref{eqn:randomPS}).

When reconstructing the coefficient functions of all remainders of the
generating sets (\ref{eqn:RallPlus}), (\ref{eqn:RsingleMinus}) and
(\ref{eqn:Rmhv}), we find it most effective to reconstruct remainder
functions in bunches of the same spinor weight. In fact, the
permutations of momenta chosen within the above 3 sets are such that
all remainders within a given set have the same spinor weight. This is
useful because there exists overlap between the vector spaces of
rational functions of each separate remainder. Namely, the dimension
of the sum of two such vector spaces is often smaller than the sum of
their dimensions. Furthermore, we find it most effective to
reconstruct denominators of lower mass dimension first.  We also
observe that permuting the momentum assignments of the functions
$\tilde r_i$ that yield the same spinor weight often yields linearly
independent coefficient functions.  We thus, after each newly
reconstructed coefficient function, ensure that the set of functions
is closed under all momentum permutations that leave the little-group
weights unchanged.  This ensures that information is recycled among
different partials, and that the symmetries of amplitudes are
exploited. Numerically, the closure of the space of functions is
conveniently checked by adding all permutations and evaluating them on
the random phase-space points (\ref{eqn:randomPS}).  We then filter
out redundant ones via Gaussian elimination.

After the reconstruction of the function coefficients is complete, we simplify
the final results considering each of the three helicity sets individually.
In fact, we perform a further basis change 
on the vector space
obtained from the union of all remainders.  We
determine a simple basis for this space, by placing the chosen basis
functions in global, or failing that local, minima in the space of
possible denominator powers \cite{Abreu:2023bdp}. Finally, the basis functions are then partial
fractioned following the approach of refs.~\cite{DeLaurentis:2019bjh,
DeLaurentis:2022otd}. 

We display the set of all coefficient functions in \cref{sec:5g-all-plus-basis,sec:5g-single-minus-basis,sec:5g-mhv-basis}. 

\subsection{Implementation}
To perform the reconstruction, for the most complicated remainder, 
we use roughly $35,\kern-0.5mm 000$ numerical
samples. Most of these samples are points on slices needed to perform
the basis change, while around $5,\kern-0.5mm 000$ are
random phase-space points. All rational functions are fitted with a single finite
field. This is possible thanks to the size of the rational numbers
appearing in the basis, with the largest ones not exceeding 2
digits. Initially the matrices $M_{ij}$ are obtained in a finite field.
To lift the matrices $M_{ij}$ from a finite field to rational
numbers we require, on the first finite field value, as many samples
as the dimension of the vector space, while subsequent finite field
values requires roughly a factor of $5$ fewer samples for  each
iteration. In the end, a single evaluation with a finite field value
not employed in the reconstruction is used as a check.

Finally, we summarize software packages used in the analytic
reconstruction.  For convenience, we take advantage of hardware
acceleration on NVIDIA GPUs with the private code \texttt{linac}
(LINear Algebra with C{\sc uda}), which we use to solve linear systems
over a finite fields for the ansatz coefficients, and to handle the
vector spaces of rational functions. Nevertheless, given the
relatively small size of the systems, a low-level CPU implementation
may also have sufficed.
We use \texttt{lips} and \texttt{pyadic} \cite{DeLaurentis:2023qhd}
for the generation and manipulations of phase-space points defined in
terms of spinors, 
and for
numerical evaluations of spinor-helicity functions.

\section{Results} 
\label{sec:results} 

\begin{table}[t]
\renewcommand{\arraystretch}{1.2}
\centering
\begin{tabular}{*{3}{C{15ex}}}
\toprule
Gluon helicities & Vector-space dimension & Generating set size \\
\colrule
$+++\;++$ & 24 & 3 \\
$+++\;+-$ & 440 & 33 \\
$+++\;--$ & 937 & 115 \\
\botrule
\end{tabular}
\caption{
  \label{tab:VS-sizes} For each helicity configuration, this table
  shows the dimension of the vector space of rational functions, and
  the number of functions in the generating set that spans the space
  upon closure under the symmetries of the helicity vector. }
\end{table}
\vspace{-2mm}

We express remainders in terms of three rational-function bases. The
function bases are obtained from a generating set of spinor-helicity
functions and symmetry operations. We denote the generating functions
as $\tilde{r}^{h_4h_5}_i$, with $h_4$ and $h_5$ labeling the three
helicity configuration: 
all-plus ($\tilde r^{++}_i$), 
single-minus ($\tilde r^{+-}_i$), 
and MHV ($\tilde r^{--}_i$). The
respective basis functions are given in \cref{sec:5g-all-plus-basis,sec:5g-single-minus-basis,sec:5g-mhv-basis}. 
The functions themselves are
expressed in terms of spinor-helicity variables, and symmetry
operations. The latter take the form
\begin{equation}
(12345\; \rightarrow \; \pm\sigma_1\sigma_2\sigma_3\sigma_4\sigma_5)
\end{equation}
where ``$-$'' denotes that the expression with the permuted labels should be
subtracted. For example,
\begin{align}
&r(1,2,3,4,5) + (12345\; \rightarrow 45321 ) = \\
& r(1,2,3,4,5) + r(4,5,3,2,1)\,,\\ 
&r(1,2,3,4,5) + (12345\; \rightarrow -45321 ) = \\
& r(1,2,3,4,5) - r(4,5,3,2,1)\,.
\end{align}
Within a rational function, the employed
convention is to apply the (anti-)symmetrization to all terms
preceding the mapping \cite[section 4.3]{DeLaurentis:2019bjh}.

\begin{table*}[ht]
\renewcommand{\arraystretch}{1.6}
\centering
\begin{tabular}{c @{\hskip 5mm} c @{\hskip 5mm} c @{\hskip 5mm} c}
\toprule
Helicity & \multirow{2}{*}{\(\text{dim}(\text{VS}(\mathcal{R}))\)} & \multicolumn{2}{c}{LCD ansatz size}  \\
remainder & & before basis change & after basis change \\
\colrule
\(R^{(2),(2, 0)}_{+++--}\) & $31$ &  $21,\kern-0.5mm 910$ & N/A\\
\(R^{(2),(2, 0)}_{++-+-}\) & $54$ &  $54,\kern-0.5mm 148$ & N/A\\
\(R^{(2),(1, 0)}_{+++--}\) & $274$ &  $163,\kern-0.5mm 635$ & $14,\kern-0.5mm 093$ \\
\(R^{(2),(1, 0)}_{+-++-}\) & $270$ & $241,\kern-0.5mm 156$ & $14,\kern-0.5mm 552$ \\
\(R^{(2),(1, 0)}_{--+++}\) & $203$ & $82,\kern-0.5mm 180$ & $25,\kern-0.5mm 620$ \\
\colrule
\(R^{(2),(1, 1)}_{+++--}\) & $31$ & $21,\kern-0.5mm 910$ & N/A\\
\(R^{(2),(1, 1)}_{++-+-}\) & $54$ & $54,\kern-0.5mm 148$ & N/A \\
\(R^{(2),(0, 1)}_{+++--}\) & $226$ & $118,\kern-0.5mm 880$ & $4,\kern-0.5mm 108$ \\
\(R^{(2),(0, 1)}_{+-++-}\) & $240$ & $209,\kern-0.5mm 018$ & N/A \\
\(R^{(2),(0, 1)}_{--+++}\) & $157$ & $76,\kern-0.5mm 845$ & $8,\kern-0.5mm 840$ \\
\(R^{(2),(-1,1)}_{+++--}\) & $25$ & $5,\kern-0.5mm 320$ & N/A \\
\(R^{(2),(-1,1)}_{++-+-}\) & $35$ & $9,\kern-0.5mm 384$ & N/A \\
\botrule
\end{tabular}
\caption{\label{tab:ansatze-sizes}
For each partial amplitude, this table shows the dimension of the
vector space of rational functions, and ansatz size in LCD form of the
most complicated function in the basis, before and after basis
change. In the cases denoted by N/A the basis change was not
required.}
\end{table*}

To obtain the function basis from the generating functions, the set
needs to be closed under the symmetries of the little group
weights. These are $\mathcal{S}_5(1,2,3,4,5)$,
$\mathcal{S}_4(1,2,3,4)$ and
$\mathcal{S}_3(1,2,3)\otimes \mathcal{S}_2(4,5)$, for all-plus,
single-minus and MHV configurations, respectively (here
$\mathcal{S}_n(1,...,n)$ denotes the group of permutations of the
elements $\{1,...,n\}$).  Table \ref{tab:VS-sizes} shows the
dimensions of the three vector spaces, and the number of generating
spinor-helicity expressions in the chosen basis. The dimensions of the
vector spaces are uniquely defined, in the sense that they are the
smallest spaces that are both closed under the symmetries and that
span all coefficients in the required partials. The number of
generating functions is representation dependent.

We note that the overlap between the rational-function spaces of the
different partials is significant. This can be observed by comparing
the sizes of the vector spaces in \cref{tab:ansatze-sizes}, to that of
their union, closed under
$\mathcal{S}_3(1,2,3)\otimes \mathcal{S}_2(4,5)$,
in \cref{tab:VS-sizes}, in spite of the fact that the spaces of
partials are not closed under all symmetries of the phase weights.

The remaining helicity configurations are obtained by re-assigning
momentum labels and/or parity conjugation. However, some of the
permutation of \cref{eq:partial_5g} will involve exchanges between
momenta in the initial and final state. This encompasses non-trivial
analytic continuation, which we perform following
ref.~\cite{Abreu:2021oya}. More explicitly, the
action of permutation $\sigma$ on a finite remainder $R$ is given by
\begin{align}\label{eq:permutation-on-remainder-with-possible-crossing}
\sigma \circ R &= \sigma \circ (\tilde r_i M_{ij} h_j) = (\sigma \circ \tilde r)_{i} M_{ij}Q_{jk} h_{k}  \\
&= (\sigma \circ \tilde r)_{i} M'_{ij} h_{j} \, .
\end{align}
Here we rely on the closure of the pentagon functions under permutations \cite{Chicherin:2020oor}.
The analytic continuation therefore amounts to simply obtaining a matrix $M'_{ij}$ for each required permutation.
In practice, these matrices are obtained by permuting the legs in the
master integrals only, and then re-mapping them to pentagon
functions. 

\subsection{Ancillary files}

We provide expressions for the finite remainders of all independent
partial amplitudes through two loops in the decomposition
of \cref{eq:partial_5g,eq:as-series,eq:Nc-Nf-series} in \cite{ancillaries}.
For each
helicity configuration, all-plus, single-minus and MHV, we organize
the results in terms of two global bases, valid for all partials and
crossings, in the files
\begin{enumerate}
\item \texttt{basis\_transcendental}$\,,$
\item \texttt{basis\_rational}$\,.$
\end{enumerate}
The constant matrices $M_{ij}$ of rational numbers are partial remainder, and
permutation specific. They are organized in subfolders labeling the
partial remainders from
subsection \ref{subsec:generating-partial-helicity-remainders}, with
the notation
\begin{enumerate}
\item[] \texttt{\{}$\vec h$\texttt{\}\_\{}$L$\texttt{\}L\_Nc\{}$n_c$\texttt{\}\_Nf\{}$n_f$\texttt{\}/}$\,,$
\end{enumerate}
where $\vec h$, $L$, $n_c$ and $n_f$ refer to helicities, number of
loops, number of $N_c$ powers and number of $N_f$ powers, as defined
earlier in the paper. For completeness of the ancillaries, we also
provide information about permutations and color structures of the
partials in the file \texttt{amp\_info}.

The matrices themselves are stored in the files
\begin{enumerate}
\item[] \texttt{rational\_matrix\_\{permutation\}}$\,,$
\end{enumerate}
where it is understood that this permutation has to be combined with
that of each partial as defined in
subsection \ref{subsec:generating-partial-helicity-remainders} and
in \texttt{amp\_info}. The order of the permutations matters, with
those defined in \ref{subsec:generating-partial-helicity-remainders}
taking precedence.

\subsection{Validation}

Our computation incorporates multiple internal consistency checks at
various stages. 
In constructing the finite remainders, we
ensure the cancellation of poles in the dimensional regulator
$\epsilon$ at every kinematic point. Further validation of the finite
field reconstruction occurs at an independent kinematic point, not
utilized for the reconstruction, and with a distinct value of the
prime.

We found agreement with the numerical evaluations for the helicity- and
color-summed squared remainders in the leading-color
approximation \cite{Abreu:2021oya}. Furthermore, we conduct additional
checks through independent computations in full color. 
We compared the one-loop hard functions against numerical evaluations by \texttt{BlackHat} \cite{Berger:2008sj} and
found agreement.
At two loops we performed
verification against the full-color all-plus calculation~\cite{Badger:2019djh}, 
and the validation of the evaluations provided in appendix~\ref{sec:referenceEvaluations} 
with an independent computation \cite{Comparison,Agarwal:2023suw}.

\subsection{Numerical evaluation}

We implement our analytic results for the partial helicity remainders, 
as well as for the NNLO hard function defined in \cref{eq:hard-function} in the \texttt{C++} library \texttt{FivePointAmplitudes}~\cite{FivePointAmplitudes},
which employs \texttt{PentagonFunctions++} \cite{Chicherin:2020oor,Chicherin:2021dyp,Abreu:2023rco} for numerical evaluation of the transcendental integral functions.
This allows us to ensure the stability of numerical evaluations via the rescue system developed in ref.~\cite{Abreu:2021oya}.
We achieve excellent numerical performance, with a single numerical evaluation of the two-loop hard function taking a few seconds in double precision on a personal desktop computer.
The average evaluation time per phase-space point on cluster nodes over the phase-space of ref.~\cite{Abreu:2021oya} is around 8s with the rescue system enabled.
For comparison, the average evaluation time for the leading-color contribution only is around 1.6s with the same setup.

\section{Conclusions}

In this work we computed the two-loop five-gluon helicity amplitudes in QCD.
The amplitudes are uniformly represented in terms of a basis of rational 
coefficient functions, transcendental functions and constant matrices of rational numbers
that link the two.
We achieved an unprecedented level of simplicity in the basis of
coefficient functions, by identifying a distinguished basis of functions 
based on their singularity structure.
However, it has to be noted
that significant complexity still remains in the transcendental functions 
and in the mentioned matrices of
rational numbers. This observation raises an intriguing possibility
of a more physically motivated basis of transcendental functions,
rooted in the cancellation of spurious singularities, that is, in 
the amplitudes' locality. Further exploration into this avenue remains an
intriguing direction for future research.

In terms of phenomenological applications, the significance of our
result becomes particularly evident when considering not just
three-jet production at NNLO, but also two-jet at N\textsuperscript{3}LO in hadron collisions. In fact,
the latter will require integrating the provided expressions in unresolved phase-space
configurations. We expect our compact basis of rational coefficient functions
to benefit both precision and stability of such computations.

Finally, the simplicity achieved in the present computation is not
confined to the specific channel under consideration, nor are the
techniques employed specific to five-point massless
kinematics. Analogous results for the quark channels of
three-jet production will follow shortly in a separate publication,
and preliminary studies show similar benefits in tackling processes
involving more challenging kinematics, such as five-point one-mass. 
In summary, we
believe that the form of the rational coefficient functions, which directly
benefits phenomenological studies, deserves further studies concerning
the mathematical structure of scattering amplitudes, and may open new
paths to precision predictions for multi-scale scattering processes at
particle colliders.

\begin{acknowledgements}

First and foremost, we gratefully acknowledge Ben Page for stimulating conversations 
and for collaboration on the basis-change algorithm \cite{Wjjcompact} that we
applied in this paper.
We thank Thomas Gehrmann for interesting discussions.
We thank Johannes Schlenk for discussions and comments on the manuscript.
Finally, we thank Bakul Agarwal, Federico Buccioni, Federica Devoto, Giulio
Gambuti, Andreas von Manteuffel, Lorenzo Tancredi for the numerical comparison
of the reference values which are displayed in
appendix~\ref{sec:referenceEvaluations}.
We gratefully acknowledge the computing resources provided by the Paul Scherrer
Insitut (PSI) and the University of Zurich (UZH).
V.S.\ has received funding from the European Research Council (ERC) under the
European Union's Horizon 2020 research and innovation programme grant agreement
101019620 (ERC Advanced Grant TOPUP).
G.D.L.'s work is supported in part by the U.K.\ Royal Society through
Grant URF\textbackslash R1\textbackslash 20109.
\end{acknowledgements}

\clearpage

\onecolumngrid
\appendix  
\newgeometry{left=0.6in, right=0.8in, top=0.8in, bottom=0.8in}

\section{Five-gluon all-plus basis functions}
\label{sec:5g-all-plus-basis}

\vspace{-4.5mm}

\hskip-2mm
\begin{tabular}{p{4.5cm}p{7.1cm}p{5.5cm}}
\begin{equation}\label{eq:all-plus-gluons-vs-basis-start}\begin{gathered}
\small \tilde{r}^{++}_{1} = \frac{[45]^2}{⟨12⟩⟨13⟩⟨23⟩}
\end{gathered}\nonumber
\end{equation}
&
\begin{equation}\hspace{-5mm}\label{eq:all-plus-gluons-vs-basis-middle}\begin{gathered}
\small \tilde{r}^{++}_{2} = \frac{-⟨14⟩⟨24⟩⟨25⟩[25]⟨35⟩[45]}{⟨12⟩^2⟨15⟩⟨23⟩⟨34⟩^2⟨45⟩}+\\
\small\frac{[24][35]}{⟨12⟩⟨15⟩⟨34⟩}+\\
\small(12345\; \rightarrow \; 23451)+\small(12345\; \rightarrow \; 34512)+\\
\small(12345\; \rightarrow \; 45123)+\small(12345\; \rightarrow \; 51234)\phantom{+}
\end{gathered}\nonumber
\end{equation}
&
\begin{equation}\label{eq:all-plus-gluons-vs-basis-end}\begin{gathered}
\small \tilde{r}^{++}_{3} = \frac{\text{tr}_5(1234)}{⟨12⟩⟨15⟩⟨25⟩⟨34⟩^2}+\\
\small(12345\; \rightarrow \; 12453)+\\
\small(12345\; \rightarrow \; 12534)\phantom{+}
\end{gathered}\nonumber
\end{equation}
\end{tabular}

%
\newgeometry{left=0.6in, right=0.8in, top=0.8in, bottom=0.8in}

\section{Five-gluon single-minus basis functions}
\label{sec:5g-single-minus-basis}

\vspace{-8.5mm}
\begin{tabular}{p{5.5cm}p{5.9cm}p{5.9cm}}
\begin{equation}\begin{gathered}
\small \tilde{r}^{+-}_{1} = \frac{[34]^3}{⟨12⟩^2[35][45]}
\end{gathered}\hspace{-2mm}\nonumber
\end{equation}
&
\begin{equation}\begin{gathered}
\small \tilde{r}^{+-}_{18} = \frac{-15⟨45⟩[13]⟨15⟩}{⟨13⟩⟨14⟩⟨24⟩^2}+\\
\small\frac{⟨45⟩^2[34][23]}{⟨14⟩^2⟨24⟩^2[24]}\phantom{+}
\end{gathered}\hspace{-2mm}\nonumber
\end{equation}
&
\begin{equation}\begin{gathered}
\small \tilde{r}^{+-}_{28} = \frac{⟨25⟩[12]^2[34]}{⟨12⟩[15]⟨34⟩⟨2|1\tiny{+}5|2]}+\\
\small(12345\; \rightarrow \; -43215)+\\
\small\frac{[12][14][34]}{⟨12⟩[15]⟨34⟩[45]}\phantom{+}
\end{gathered}\hspace{-2mm}\nonumber
\end{equation}
\\[-22mm]
\begin{equation}\begin{gathered}
\small \tilde{r}^{+-}_{2} = \frac{⟨45⟩[23]⟨15⟩}{⟨14⟩^3⟨23⟩}
\end{gathered}\hspace{-2mm}\nonumber
\end{equation}
&
\vspace{6mm}
\begin{equation}\begin{gathered}
\small \tilde{r}^{+-}_{19} = \frac{-3⟨23⟩⟨25⟩⟨35⟩[12]}{⟨12⟩⟨23⟩^2⟨34⟩^2}+\\
\small\frac{-2⟨25⟩^2⟨35⟩[12]}{⟨12⟩⟨23⟩^2⟨34⟩⟨45⟩}\phantom{+}
\end{gathered}\hspace{-2mm}\nonumber
\end{equation}
&
\vspace{8mm}
\begin{equation}\begin{gathered}
\small \tilde{r}^{+-}_{29} = \frac{[35]⟨45⟩^3⟨15⟩^2[45]}{⟨14⟩^4⟨23⟩⟨24⟩[34]⟨35⟩}+\\
\small\frac{-4[35]⟨45⟩^3⟨13⟩⟨15⟩}{⟨14⟩^4⟨23⟩⟨24⟩⟨35⟩}+\\
\small\frac{4⟨45⟩^3⟨12⟩[24]}{⟨14⟩^3⟨23⟩⟨24⟩⟨35⟩}\phantom{+}
\end{gathered}\hspace{-2mm}\nonumber
\end{equation}
\\[-42mm]
\begin{equation}\begin{gathered}
\small \tilde{r}^{+-}_{3} = \frac{⟨25⟩[23]⟨35⟩}{⟨14⟩^2⟨23⟩^2}
\end{gathered}\hspace{-2mm}\nonumber
\end{equation}
&
\vspace{16mm}
\begin{equation}\begin{gathered}
\small \tilde{r}^{+-}_{20} = \frac{12⟨25⟩[14]⟨15⟩}{⟨12⟩⟨14⟩⟨23⟩^2}+\\
\small\frac{⟨25⟩⟨45⟩[14][24]}{⟨12⟩[12]⟨14⟩⟨23⟩^2}\phantom{+}
\end{gathered}\hspace{-2mm}\nonumber
\end{equation}
&
\vspace{28mm}
\begin{equation}\begin{gathered}
\small \tilde{r}^{+-}_{30} = \frac{-2[12][23]⟨25⟩^2[24][34]}{⟨12⟩⟨2|1\tiny{+}5|2]^3}+\\
\small\frac{-[12]⟨25⟩⟨45⟩[24][34]}{⟨12⟩⟨34⟩⟨2|1\tiny{+}5|2]^2}+\\
\small\frac{2⟨35⟩⟨45⟩[12][34]}{⟨12⟩⟨34⟩^2⟨2|1\tiny{+}5|2]}\phantom{+}
\end{gathered}\hspace{-2mm}\nonumber
\end{equation}
\\[-60mm]
\begin{equation}\begin{gathered}
\small \tilde{r}^{+-}_{4} = \frac{[45]⟨25⟩^3}{⟨12⟩^2⟨23⟩^2⟨24⟩}
\end{gathered}\hspace{-2mm}\nonumber
\end{equation}
&
\vspace{25mm}
\begin{equation}\begin{gathered}
\small \tilde{r}^{+-}_{21} = \frac{5⟨35⟩^2[45][23]^2}{⟨13⟩^2⟨34⟩^2[34][35]}+\\
\small\frac{17⟨35⟩^2[24][23]}{⟨13⟩^2⟨34⟩^2[34]}\phantom{+}
\end{gathered}\hspace{-2mm}\nonumber
\end{equation}
&
\vspace{46mm}
\begin{equation}\begin{gathered}
\small \tilde{r}^{+-}_{31} = \frac{-2⟨35⟩⟨25⟩[34]}{⟨12⟩^2⟨23⟩⟨34⟩}+\\
\small\frac{2⟨45⟩[14][34]⟨15⟩}{⟨12⟩^2⟨34⟩⟨1|2\tiny{+}3|1]}+\\
\small\frac{3[23]⟨35⟩[13][14]⟨15⟩}{⟨14⟩⟨23⟩⟨1|2\tiny{+}3|1]^2}+\\
\small\frac{-6[12][23][13][14]⟨15⟩^2}{⟨14⟩⟨1|2\tiny{+}3|1]^3}\phantom{+}
\end{gathered}\hspace{-2mm}\nonumber
\end{equation}
\\[-88mm]
\begin{equation}\begin{gathered}
\small \tilde{r}^{+-}_{5} = \frac{⟨25⟩^2[12]}{⟨12⟩⟨23⟩⟨24⟩⟨34⟩}
\end{gathered}\hspace{-2mm}\nonumber
\end{equation}
&
\vspace{36mm}
\begin{equation}\begin{gathered}
\small \tilde{r}^{+-}_{22} = \frac{[35]⟨45⟩^3[24]⟨15⟩}{⟨14⟩^3⟨24⟩⟨34⟩[34]⟨35⟩}+\\
\small\frac{8⟨45⟩^3[24]}{⟨14⟩^2⟨24⟩⟨34⟩⟨35⟩}\phantom{+}
\end{gathered}\hspace{-2mm}\nonumber
\end{equation}
&
\vspace{75mm}
\begin{equation}\hspace{-2mm}\begin{gathered}
\small \tilde{r}^{+-}_{32} = \frac{[12]⟨12⟩[23]⟨45⟩[14]⟨15⟩}{⟨14⟩^2⟨23⟩⟨1|2\tiny{+}3|1]^2}+\\
\small\frac{-⟨12⟩[23]⟨35⟩⟨45⟩⟨15⟩[15]}{⟨14⟩^3⟨23⟩^2⟨1|2\tiny{+}3|1]}+\\
\small\frac{2[12]⟨12⟩[23]⟨45⟩⟨15⟩}{⟨14⟩^3⟨23⟩⟨1|2\tiny{+}3|1]}+\\
\small\frac{-⟨35⟩[35]⟨45⟩⟨15⟩}{⟨14⟩^3⟨23⟩^2}+\\
\small\frac{⟨25⟩⟨45⟩[24]}{⟨14⟩^2⟨23⟩^2}\phantom{+}
\end{gathered}\hspace{-10mm}\nonumber
\end{equation}
\\[-128mm]
\begin{equation}\begin{gathered}
\small \tilde{r}^{+-}_{6} = \frac{[25]⟨25⟩^3}{⟨12⟩^2⟨23⟩⟨24⟩⟨34⟩}
\end{gathered}\hspace{-2mm}\nonumber
\end{equation}
&
\vspace{47mm}
\begin{equation}\begin{gathered}
\small \tilde{r}^{+-}_{23} = \frac{-⟨35⟩⟨25⟩^2[12][45]}{⟨12⟩⟨23⟩^3[25]⟨45⟩}+\\
\small\frac{-10⟨35⟩⟨25⟩^2[12]}{⟨12⟩⟨23⟩^2⟨34⟩⟨45⟩}\phantom{+}
\end{gathered}\hspace{-2mm}\nonumber
\end{equation}
&
\vspace{116mm}
\begin{equation}\begin{gathered}
\small \tilde{r}^{+-}_{33} = \frac{[24][14][13]⟨15⟩}{⟨14⟩⟨23⟩[45]⟨1|2\tiny{+}3|1]}+\\
\small\frac{-⟨25⟩[12][24][13]}{⟨12⟩[15]⟨34⟩⟨2|1\tiny{+}5|2]}+\\
\small\frac{2[12][14][13]⟨15⟩^2}{⟨14⟩⟨23⟩⟨1|2\tiny{+}3|1]^2}+\\
\small\frac{-2⟨25⟩^2[12][24][23]}{⟨12⟩⟨34⟩⟨2|1\tiny{+}5|2]^2}+\\
\small(12345\; \rightarrow \; -43215)+\\
\small\frac{-⟨24⟩[13]⟨13⟩[14][24]}{⟨12⟩⟨14⟩[15]⟨23⟩⟨34⟩[45]}\phantom{+}
\end{gathered}\hspace{-10mm}\nonumber
\end{equation}
\\[-172mm]
\begin{equation}\begin{gathered}
\small \tilde{r}^{+-}_{7} = \frac{⟨24⟩⟨25⟩⟨13⟩⟨15⟩[34]}{⟨12⟩^4⟨34⟩^2}
\end{gathered}\hspace{-2mm}\nonumber
\end{equation}
&
\vspace{57mm}
\begin{equation}\begin{gathered}
\small \tilde{r}^{+-}_{24} = \frac{3⟨34⟩⟨35⟩⟨25⟩^2[14][24]}{[12]⟨14⟩⟨15⟩⟨23⟩^4}+\\
\small\frac{⟨35⟩⟨45⟩[14][24]}{[12]⟨13⟩⟨14⟩⟨23⟩^2}\phantom{+}
\end{gathered}\hspace{-2mm}\nonumber
\end{equation}
&
\\[-8.5mm]
\vspace{-73mm}
\begin{equation}\begin{gathered}
\small \tilde{r}^{+-}_{8} = \frac{[15]⟨35⟩^2⟨25⟩^2}{⟨15⟩⟨23⟩^3⟨24⟩⟨34⟩}
\end{gathered}\hspace{-2mm}\nonumber
\end{equation}
&
\begin{equation}\begin{gathered}
\small \tilde{r}^{+-}_{25} = \frac{⟨34⟩⟨35⟩⟨25⟩[14][34]}{⟨13⟩[13]⟨14⟩⟨23⟩^3}+\\
\small\frac{-5⟨35⟩⟨45⟩[14]^2[23]}{[12]⟨13⟩[13]⟨14⟩⟨23⟩^2}\phantom{+}
\end{gathered}\hspace{-2mm}\nonumber
\end{equation}
&
\\[-8.5mm]
\vspace{-82mm}
\begin{equation}\begin{gathered}
\small \tilde{r}^{+-}_{9} = \frac{[23]⟨45⟩[34]⟨25⟩}{⟨12⟩⟨14⟩⟨24⟩^2[24]}
\end{gathered}\hspace{-2mm}\nonumber
\end{equation}
&
\begin{equation}\begin{gathered}
\small \tilde{r}^{+-}_{26} = \frac{5⟨34⟩⟨35⟩[45]⟨25⟩^3[24]}{⟨12⟩⟨15⟩⟨23⟩^4[25]⟨45⟩}+\\
\small\frac{4⟨35⟩⟨25⟩^2[12][45]}{⟨12⟩⟨23⟩^3[25]⟨45⟩}\phantom{+}
\end{gathered}\hspace{-2mm}\nonumber
\end{equation}
&
\\[-8.5mm]
\vspace{-93mm}
\begin{equation}\begin{gathered}
\small \tilde{r}^{+-}_{10} = \frac{⟨45⟩[14][23]⟨15⟩}{⟨14⟩^2⟨23⟩⟨1|2\tiny{+}3|1]}
\end{gathered}\hspace{-2mm}\nonumber
\end{equation}
&
\begin{equation}\begin{gathered}
\small \tilde{r}^{+-}_{27} = \frac{-[45]⟨15⟩⟨25⟩^3[12]}{⟨12⟩^2⟨23⟩^2⟨45⟩⟨2|1\tiny{+}5|2]}+\\
\small\frac{-3[34]⟨25⟩⟨45⟩⟨35⟩[25]}{⟨12⟩^2⟨34⟩^2⟨2|1\tiny{+}5|2]}\phantom{+}
\end{gathered}\hspace{-2mm}\nonumber
\end{equation}
&
\\[-113mm]
\begin{equation}\begin{gathered}
\small \tilde{r}^{+-}_{11} = \frac{[12][34]⟨25⟩^2}{⟨12⟩⟨23⟩⟨24⟩⟨2|1\tiny{+}5|2]}
\end{gathered}\hspace{-2mm}\nonumber
\end{equation}
&
&
\\[-8.5mm]
\begin{equation}\begin{gathered}
\small \tilde{r}^{+-}_{12} = \frac{⟨25⟩^2⟨45⟩[34][23]}{⟨12⟩⟨15⟩⟨24⟩^3[24]}
\end{gathered}\hspace{-2mm}\nonumber
\end{equation}
&
&
\\[-8.5mm]
\begin{equation}\begin{gathered}
\small \tilde{r}^{+-}_{13} = \frac{⟨35⟩^2⟨25⟩[25][14]}{[12]⟨13⟩⟨14⟩⟨23⟩^3}
\end{gathered}\hspace{-2mm}\nonumber
\end{equation}
&
&
\\[-8.5mm]
\begin{equation}\begin{gathered}
\small \tilde{r}^{+-}_{14} = \frac{⟨45⟩[14]^2[23]⟨15⟩}{⟨14⟩⟨23⟩⟨1|2\tiny{+}3|1]^2}
\end{gathered}\hspace{-2mm}\nonumber
\end{equation}
&
&
\\[-8.5mm]
\begin{equation}\begin{gathered}
\small \tilde{r}^{+-}_{15} = \frac{⟨35⟩^2⟨25⟩[25][34]}{⟨12⟩⟨13⟩⟨23⟩^2[23]⟨34⟩}
\end{gathered}\hspace{-2mm}\nonumber
\end{equation}
&
&
\\[-8.5mm]
\begin{equation}\begin{gathered}
\small \tilde{r}^{+-}_{16} = \frac{[14][13]⟨15⟩^2}{⟨12⟩^2⟨34⟩⟨1|2\tiny{+}5|1]}+\\
\small(12345\; \rightarrow \; -21435)\phantom{+}
\end{gathered}\hspace{-2mm}\nonumber
\end{equation}
&
&
\\[-8.5mm]
\begin{equation}\begin{gathered}
\small \tilde{r}^{+-}_{17} = \frac{⟨35⟩⟨45⟩[34]^2}{⟨12⟩^2⟨34⟩⟨3|1\tiny{+}2|3]}+\\
\small(12345\; \rightarrow \; -21435)\phantom{+}
\end{gathered}\hspace{-2mm}\nonumber
\end{equation}
&
&
\\[-8.5mm]
\end{tabular}

\newgeometry{left=0.6in, right=0.8in, top=0.8in, bottom=0.8in}

\section{Five-gluon MHV basis functions}
\label{sec:5g-mhv-basis}

\vspace{-8.5mm}
\hskip+0.0cm
\begin{tabular}{p{5.0cm}p{5.9cm}p{5.9cm}}
\begin{equation}\begin{gathered}
\small \tilde{r}^{--}_{1} = \frac{⟨45⟩^2}{⟨12⟩⟨13⟩⟨23⟩}
\end{gathered}\nonumber
\end{equation}
&
\begin{equation}\begin{gathered}
\small \tilde{r}^{--}_{20} = \frac{[13]^2⟨45⟩}{⟨23⟩⟨25⟩[35][45]}
\end{gathered}\nonumber
\end{equation}
&
\begin{equation}\begin{gathered}
\small \tilde{r}^{--}_{39} = \frac{⟨25⟩^3[23]^2}{⟨12⟩^2⟨23⟩[34]^2⟨35⟩}
\end{gathered}\nonumber
\end{equation}
\\[-8.5mm]
\begin{equation}\begin{gathered}
\small \tilde{r}^{--}_{2} = \frac{⟨45⟩^3}{⟨12⟩^2⟨34⟩⟨35⟩}
\end{gathered}\nonumber
\end{equation}
&
\begin{equation}\begin{gathered}
\small \tilde{r}^{--}_{21} = \frac{[23]⟨45⟩^2}{⟨12⟩⟨13⟩⟨35⟩[35]}
\end{gathered}\nonumber
\end{equation}
&
\begin{equation}\begin{gathered}
\small \tilde{r}^{--}_{40} = \frac{⟨25⟩^3[25]^2}{⟨12⟩^2⟨23⟩⟨35⟩[45]^2}
\end{gathered}\nonumber
\end{equation}
\\[-8.5mm]
\begin{equation}\begin{gathered}
\small \tilde{r}^{--}_{3} = \frac{⟨45⟩^3}{⟨12⟩⟨15⟩⟨23⟩⟨34⟩}
\end{gathered}\nonumber
\end{equation}
&
\begin{equation}\begin{gathered}
\small \tilde{r}^{--}_{22} = \frac{[13][23]^2}{⟨12⟩[25][34][45]}
\end{gathered}\nonumber
\end{equation}
&
\begin{equation}\begin{gathered}
\small \tilde{r}^{--}_{41} = \frac{[12]⟨35⟩⟨15⟩⟨14⟩}{⟨12⟩⟨13⟩^3[14]}
\end{gathered}\nonumber
\end{equation}
\\[-8.5mm]
\begin{equation}\begin{gathered}
\small \tilde{r}^{--}_{4} = \frac{[14][12][35]}{⟨23⟩[45]^3}
\end{gathered}\nonumber
\end{equation}
&
\begin{equation}\begin{gathered}
\small \tilde{r}^{--}_{23} = \frac{[12]^2⟨45⟩}{⟨13⟩[15]⟨23⟩[24]}
\end{gathered}\nonumber
\end{equation}
&
\begin{equation}\begin{gathered}
\small \tilde{r}^{--}_{42} = \frac{⟨45⟩^3[23]}{⟨14⟩⟨15⟩⟨23⟩⟨24⟩[24]}
\end{gathered}\nonumber
\end{equation}
\\[-8.5mm]
\begin{equation}\begin{gathered}
\small \tilde{r}^{--}_{5} = \frac{⟨45⟩^2⟨24⟩}{⟨12⟩^2⟨23⟩⟨34⟩}
\end{gathered}\nonumber
\end{equation}
&
\begin{equation}\begin{gathered}
\small \tilde{r}^{--}_{24} = \frac{⟨25⟩⟨34⟩^2[12]}{⟨13⟩⟨23⟩^3[25]}
\end{gathered}\nonumber
\end{equation}
&
\begin{equation}\begin{gathered}
\small \tilde{r}^{--}_{43} = \frac{[35][15]⟨35⟩⟨25⟩}{⟨12⟩⟨23⟩^2[45]^2}
\end{gathered}\nonumber
\end{equation}
\\[-8.5mm]
\begin{equation}\begin{gathered}
\small \tilde{r}^{--}_{6} = \frac{⟨15⟩⟨14⟩⟨45⟩}{⟨12⟩^2⟨13⟩^2}
\end{gathered}\nonumber
\end{equation}
&
\begin{equation}\begin{gathered}
\small \tilde{r}^{--}_{25} = \frac{⟨25⟩[14][25]^2}{⟨13⟩⟨23⟩[45]^3}
\end{gathered}\nonumber
\end{equation}
&
\begin{equation}\begin{gathered}
\small \tilde{r}^{--}_{44} = \frac{[25]⟨25⟩^2[13]}{⟨12⟩^2[14]⟨23⟩[45]}
\end{gathered}\nonumber
\end{equation}
\\[-8.5mm]
\begin{equation}\begin{gathered}
\small \tilde{r}^{--}_{7} = \frac{[12]^2⟨45⟩}{⟨34⟩⟨35⟩[45]^2}
\end{gathered}\nonumber
\end{equation}
&
\begin{equation}\begin{gathered}
\small \tilde{r}^{--}_{26} = \frac{[13]⟨45⟩^3}{⟨13⟩[14]⟨15⟩⟨24⟩^2}
\end{gathered}\nonumber
\end{equation}
&
\begin{equation}\begin{gathered}
\small \tilde{r}^{--}_{45} = \frac{[12]⟨45⟩^2[13]}{⟨12⟩⟨14⟩[14]^2⟨34⟩}
\end{gathered}\nonumber
\end{equation}
\\[-8.5mm]
\begin{equation}\begin{gathered}
\small \tilde{r}^{--}_{8} = \frac{[25][14]^2[35]}{⟨23⟩[45]^4}
\end{gathered}\nonumber
\end{equation}
&
\begin{equation}\begin{gathered}
\small \tilde{r}^{--}_{27} = \frac{[13]^3[25]}{⟨12⟩[15]^2[34][45]}
\end{gathered}\nonumber
\end{equation}
&
\begin{equation}\begin{gathered}
\small \tilde{r}^{--}_{46} = \frac{⟨14⟩[34]⟨45⟩^2}{⟨12⟩^2⟨13⟩⟨24⟩[24]}
\end{gathered}\nonumber
\end{equation}
\\[-8.5mm]
\begin{equation}\begin{gathered}
\small \tilde{r}^{--}_{9} = \frac{[23]^2⟨34⟩}{⟨13⟩⟨14⟩[45]^2}
\end{gathered}\nonumber
\end{equation}
&
\begin{equation}\begin{gathered}
\small \tilde{r}^{--}_{28} = \frac{[12]⟨45⟩^3}{⟨14⟩⟨24⟩⟨35⟩^2[45]}
\end{gathered}\nonumber
\end{equation}
&
\begin{equation}\begin{gathered}
\small \tilde{r}^{--}_{47} = \frac{⟨34⟩[13]^2⟨45⟩}{⟨23⟩^2[35]⟨3|1\tiny{+}5|3]}
\end{gathered}\nonumber
\end{equation}
\\[-8.5mm]
\begin{equation}\begin{gathered}
\small \tilde{r}^{--}_{10} = \frac{[13]^2⟨34⟩⟨24⟩}{⟨23⟩^3[35]^2}
\end{gathered}\nonumber
\end{equation}
&
\begin{equation}\begin{gathered}
\small \tilde{r}^{--}_{29} = \frac{⟨25⟩⟨14⟩^2⟨24⟩⟨45⟩}{⟨12⟩^4⟨34⟩^2}
\end{gathered}\nonumber
\end{equation}
&
\begin{equation}\begin{gathered}
\small \tilde{r}^{--}_{48} = \frac{[13]⟨14⟩^2⟨15⟩[25]}{⟨12⟩⟨13⟩^3[35]^2}
\end{gathered}\nonumber
\end{equation}
\\[-8.5mm]
\begin{equation}\begin{gathered}
\small \tilde{r}^{--}_{11} = \frac{⟨34⟩⟨14⟩⟨45⟩^2}{⟨13⟩^3⟨24⟩^2}
\end{gathered}\nonumber
\end{equation}
&
\begin{equation}\begin{gathered}
\small \tilde{r}^{--}_{30} = \frac{[14]⟨15⟩⟨14⟩^2}{⟨12⟩^2⟨13⟩^2[45]}
\end{gathered}\nonumber
\end{equation}
&
\begin{equation}\begin{gathered}
\small \tilde{r}^{--}_{49} = \frac{[12]^2[23]^2⟨45⟩}{[24][25]⟨2|1\tiny{+}5|2]^2}
\end{gathered}\nonumber
\end{equation}
\\[-8.5mm]
\begin{equation}\begin{gathered}
\small \tilde{r}^{--}_{12} = \frac{⟨34⟩⟨14⟩⟨35⟩^2}{⟨13⟩^3⟨23⟩^2}
\end{gathered}\nonumber
\end{equation}
&
\begin{equation}\begin{gathered}
\small \tilde{r}^{--}_{31} = \frac{⟨34⟩[12]^2⟨24⟩}{⟨13⟩^2[15]^2⟨23⟩}
\end{gathered}\nonumber
\end{equation}
&
\begin{equation}\begin{gathered}
\small \tilde{r}^{--}_{50} = \frac{⟨24⟩^2[12]^2⟨35⟩}{⟨23⟩^3[25]⟨2|1\tiny{+}5|2]}
\end{gathered}\nonumber
\end{equation}
\\[-8.5mm]
\begin{equation}\begin{gathered}
\small \tilde{r}^{--}_{13} = \frac{⟨35⟩^3⟨14⟩^2}{⟨13⟩^4⟨23⟩⟨25⟩}
\end{gathered}\nonumber
\end{equation}
&
\begin{equation}\begin{gathered}
\small \tilde{r}^{--}_{32} = \frac{⟨35⟩⟨25⟩[23]^2}{⟨12⟩^2⟨23⟩[24]^2}
\end{gathered}\nonumber
\end{equation}
&
\begin{equation}\begin{gathered}
\small \tilde{r}^{--}_{51} = \frac{⟨45⟩^2[13]^2⟨14⟩}{⟨13⟩⟨25⟩^2⟨34⟩[35]^2}
\end{gathered}\nonumber
\end{equation}
\\[-8.5mm]
\begin{equation}\begin{gathered}
\small \tilde{r}^{--}_{14} = \frac{[12][23]⟨14⟩}{⟨13⟩^2[35][45]}
\end{gathered}\nonumber
\end{equation}
&
\begin{equation}\begin{gathered}
\small \tilde{r}^{--}_{33} = \frac{[13]⟨34⟩⟨35⟩^2}{⟨13⟩^2[14]⟨23⟩^2}
\end{gathered}\nonumber
\end{equation}
&
\begin{equation}\begin{gathered}
\small \tilde{r}^{--}_{52} = \frac{[12]^2⟨45⟩^2⟨23⟩}{⟨12⟩⟨13⟩[14]^2⟨34⟩^2}
\end{gathered}\nonumber
\end{equation}
\\[-8.5mm]
\begin{equation}\begin{gathered}
\small \tilde{r}^{--}_{15} = \frac{⟨25⟩⟨45⟩[13]}{⟨12⟩^2[14]⟨23⟩}
\end{gathered}\nonumber
\end{equation}
&
\begin{equation}\begin{gathered}
\small \tilde{r}^{--}_{34} = \frac{⟨45⟩^2[23]^2}{⟨12⟩⟨15⟩⟨25⟩[25]^2}
\end{gathered}\nonumber
\end{equation}
&
\begin{equation}\begin{gathered}
\small \tilde{r}^{--}_{53} = \frac{⟨45⟩^2⟨15⟩[25]^2}{⟨13⟩⟨14⟩^2⟨35⟩[45]^2}
\end{gathered}\nonumber
\end{equation}
\\[-8.5mm]
\begin{equation}\begin{gathered}
\small \tilde{r}^{--}_{16} = \frac{[15]⟨45⟩⟨25⟩}{⟨12⟩⟨23⟩^2[45]}
\end{gathered}\nonumber
\end{equation}
&
\begin{equation}\begin{gathered}
\small \tilde{r}^{--}_{35} = \frac{[13]^2⟨45⟩^2}{⟨13⟩[14]⟨24⟩^2[34]}
\end{gathered}\nonumber
\end{equation}
&
\begin{equation}\begin{gathered}
\small \tilde{r}^{--}_{54} = \frac{[12]^2⟨35⟩[23]^3}{[24]^2[25]⟨2|1\tiny{+}5|2]^2}
\end{gathered}\nonumber
\end{equation}
\\[-8.5mm]
\begin{equation}\begin{gathered}
\small \tilde{r}^{--}_{17} = \frac{⟨35⟩[13]⟨24⟩}{⟨12⟩⟨23⟩^2[45]}
\end{gathered}\nonumber
\end{equation}
&
\begin{equation}\begin{gathered}
\small \tilde{r}^{--}_{36} = \frac{⟨14⟩[13]^2⟨45⟩}{⟨12⟩^2⟨15⟩[15]^2}
\end{gathered}\nonumber
\end{equation}
&
\begin{equation}\begin{gathered}
\small \tilde{r}^{--}_{55} = \frac{⟨34⟩[23]⟨45⟩^2}{⟨13⟩⟨15⟩⟨23⟩⟨24⟩[25]}
\end{gathered}\nonumber
\end{equation}
\\[-8.5mm]
\begin{equation}\begin{gathered}
\small \tilde{r}^{--}_{18} = \frac{[12]⟨24⟩⟨45⟩}{⟨12⟩⟨23⟩^2[25]}
\end{gathered}\nonumber
\end{equation}
&
\begin{equation}\begin{gathered}
\small \tilde{r}^{--}_{37} = \frac{[34]⟨34⟩^2⟨35⟩^2}{⟨13⟩^3⟨23⟩^2[14]}
\end{gathered}\nonumber
\end{equation}
&
\begin{equation}\begin{gathered}
\small \tilde{r}^{--}_{56} = \frac{[23][13]⟨45⟩^2}{⟨15⟩⟨23⟩⟨25⟩[25][35]}
\end{gathered}\nonumber
\end{equation}
\\[-8.5mm]
\begin{equation}\begin{gathered}
\small \tilde{r}^{--}_{19} = \frac{[13]^2[23]}{⟨23⟩[34][35][45]}
\end{gathered}\nonumber
\end{equation}
&
\begin{equation}\begin{gathered}
\small \tilde{r}^{--}_{38} = \frac{⟨45⟩⟨24⟩^2⟨35⟩⟨14⟩^2}{⟨12⟩^4⟨34⟩^3}
\end{gathered}\nonumber
\end{equation}
&
\begin{equation}\begin{gathered}
\small \tilde{r}^{--}_{57} = \frac{⟨25⟩[25]⟨45⟩^2}{⟨13⟩⟨15⟩⟨23⟩⟨24⟩[45]}
\end{gathered}\nonumber
\end{equation}
\end{tabular}

\pagebreak

\begin{tabular}{p{5.5cm}p{5.9cm}p{5.9cm}}
\hskip-0.2cm
\begin{equation}\hspace{-2mm}\begin{gathered}
\small \tilde{r}^{--}_{58} = \frac{[12]⟨45⟩^2[23]}{⟨12⟩[24]⟨34⟩⟨2|1\tiny{+}5|2]}
\end{gathered}\hspace{-2mm}\nonumber
\end{equation}
&
\begin{equation}\begin{gathered}
\small \tilde{r}^{--}_{77} = \frac{⟨34⟩⟨45⟩[35]⟨35⟩^2}{⟨13⟩^2⟨23⟩^2⟨3|1\tiny{+}2|3]}
\end{gathered}\nonumber
\end{equation}
&
\begin{equation}\begin{gathered}
\small \tilde{r}^{--}_{92} = \frac{[23]^2⟨25⟩⟨24⟩^2}{⟨12⟩^2⟨23⟩⟨35⟩[35]^2}+\\
\small \quad \frac{2⟨14⟩[23]⟨25⟩^2⟨24⟩}{⟨12⟩^3⟨23⟩⟨35⟩[35]}
\end{gathered}\hspace{-2mm}\nonumber
\end{equation}
\\[-20mm]
\begin{equation}\begin{gathered}
\small \tilde{r}^{--}_{59} = \frac{[12]^2[13]⟨45⟩}{⟨13⟩[14][15]⟨1|2\tiny{+}3|1]}
\end{gathered}\nonumber
\end{equation}
&
\begin{equation}\begin{gathered}
\small \tilde{r}^{--}_{78} = \frac{⟨34⟩[12][23]^3⟨25⟩^2}{⟨12⟩[24]⟨2|1\tiny{+}5|2]^3}
\end{gathered}\nonumber
\end{equation}
&
\vspace{8mm}
\begin{equation}\begin{gathered}
\small \tilde{r}^{--}_{93} = \frac{2⟨25⟩^3[25]⟨14⟩⟨24⟩}{⟨12⟩^3⟨23⟩^2⟨35⟩[35]}+\\
\small \quad \frac{-⟨25⟩^3[25]^2⟨24⟩^2}{⟨12⟩^2⟨23⟩^3⟨35⟩[35]^2}
\end{gathered}\hspace{-2mm}\nonumber
\end{equation}
\\[-31mm]
\begin{equation}\begin{gathered}
\small \tilde{r}^{--}_{60} = \frac{[34][12][35]⟨45⟩}{⟨12⟩[45]^2⟨5|1\tiny{+}2|5]}
\end{gathered}\nonumber
\end{equation}
&
\begin{equation}\begin{gathered}
\small \tilde{r}^{--}_{79} = \frac{⟨45⟩^2⟨34⟩^2[34]^2}{⟨12⟩⟨13⟩⟨23⟩⟨35⟩^2[35]^2}
\end{gathered}\nonumber
\end{equation}
&
\vspace{19mm}
\begin{equation}\begin{gathered}
\small \tilde{r}^{--}_{94} = \frac{-[12][23]^2⟨14⟩⟨23⟩}{⟨13⟩^3[13][35][45]}+\\
\small \quad \frac{[12]^2⟨15⟩[23]⟨23⟩}{⟨13⟩^3[13][14][45]}\phantom{+}
\end{gathered}\hspace{-2mm}\nonumber
\end{equation}
\\[-42mm]
\begin{equation}\begin{gathered}
\small \tilde{r}^{--}_{61} = \frac{[13]⟨34⟩⟨24⟩[12]^2}{[15]⟨23⟩^3[23][25]}
\end{gathered}\nonumber
\end{equation}
&
\begin{equation}\begin{gathered}
\small \tilde{r}^{--}_{80} = \frac{⟨45⟩^2⟨35⟩^2[35]⟨25⟩}{⟨15⟩^2⟨23⟩^3⟨5|1\tiny{+}4|5]}
\end{gathered}\nonumber
\end{equation}
&
\vspace{30mm}
\begin{equation}\begin{gathered}
\small \tilde{r}^{--}_{95} = \frac{2[25][15]⟨45⟩^3[34]^2}{⟨12⟩[45]⟨5|1\tiny{+}2|5]^3}+\\
\small \quad\; \frac{3[35]⟨15⟩[12][13]⟨45⟩}{⟨12⟩[45]⟨5|1\tiny{+}2|5]^2}\phantom{+}
\end{gathered}\hspace{-2mm}\nonumber
\end{equation}
\\[-51mm]
\begin{equation}\begin{gathered}
\small \tilde{r}^{--}_{62} = \frac{[23]⟨45⟩⟨35⟩⟨34⟩}{⟨13⟩^2⟨23⟩⟨3|1\tiny{+}2|3]}
\end{gathered}\nonumber
\end{equation}
&
\begin{equation}\begin{gathered}
\small \tilde{r}^{--}_{81} = \frac{[12]⟨45⟩[34]⟨34⟩}{⟨13⟩⟨23⟩⟨35⟩[35][45]}
\end{gathered}\nonumber
\end{equation}
&
\vspace{40mm}
\begin{equation}\begin{gathered}
\small \tilde{r}^{--}_{96} = \frac{2[34][14][35][25]⟨45⟩}{⟨12⟩[45]^3⟨5|1\tiny{+}2|5]}+\\
\small\frac{[34]^2[15][25]⟨45⟩^2}{⟨12⟩[45]^2⟨5|1\tiny{+}2|5]^2}\phantom{+}
\end{gathered}\nonumber
\end{equation}
\\[-62mm]
\begin{equation}\begin{gathered}
\small \tilde{r}^{--}_{63} = \frac{[12]⟨25⟩[23]^2⟨35⟩}{⟨12⟩⟨23⟩⟨24⟩[24]^3}
\end{gathered}\nonumber
\end{equation}
&
\begin{equation}\begin{gathered}
\small \tilde{r}^{--}_{82} = \frac{[12]⟨35⟩⟨45⟩^3[25]}{⟨12⟩[24]^2⟨25⟩⟨34⟩^3}
\end{gathered}\nonumber
\end{equation}
&
\vspace{50mm}
\begin{equation}\hspace{-2mm}\begin{gathered}
\small \tilde{r}^{--}_{97} = \frac{-⟨34⟩^2⟨45⟩[34][12]}{⟨13⟩⟨14⟩[15]⟨23⟩⟨35⟩[45]}+\\
\small\frac{2⟨45⟩^3[14][13]}{⟨14⟩[15]⟨25⟩^2⟨35⟩[45]}\phantom{+}
\end{gathered}\hspace{-3mm}\nonumber
\end{equation}
\\[-73mm]
\begin{equation}\begin{gathered}
\small \tilde{r}^{--}_{64} = \frac{⟨24⟩[23]⟨25⟩^2⟨15⟩}{⟨12⟩^3⟨23⟩[34]⟨35⟩}
\end{gathered}\nonumber
\end{equation}
&
\begin{equation}\begin{gathered}
\small \tilde{r}^{--}_{83} = \frac{[12][13]⟨45⟩⟨25⟩^2}{⟨12⟩[14]^2⟨15⟩⟨23⟩⟨24⟩}
\end{gathered}\nonumber
\end{equation}
&
\vspace{65mm}
\begin{equation}\begin{gathered}
\small \tilde{r}^{--}_{98} = \frac{[14][15]⟨45⟩⟨24⟩[24]}{⟨23⟩^2[45]^2⟨4|1\tiny{+}5|4]}+\\
\small\frac{-[14][15][25]⟨25⟩⟨45⟩}{⟨23⟩^2[45]^2⟨5|1\tiny{+}4|5]}\phantom{+}
\end{gathered}\nonumber
\end{equation}
\\[-87mm]
\begin{equation}\begin{gathered}
\small \tilde{r}^{--}_{65} = \frac{[35]^2⟨25⟩⟨15⟩⟨45⟩}{⟨12⟩^3[45]⟨5|1\tiny{+}2|5]}
\end{gathered}\nonumber
\end{equation}
&
\begin{equation}\begin{gathered}
\small \tilde{r}^{--}_{84} = \frac{⟨25⟩^2⟨24⟩[23]^2[12]}{⟨12⟩⟨23⟩[34]⟨2|1\tiny{+}5|2]^2}
\end{gathered}\nonumber
\end{equation}
&
\vspace{75mm}
\begin{equation}\begin{gathered}
\small \tilde{r}^{--}_{99} = \frac{-[12]⟨24⟩[34]⟨45⟩}{⟨12⟩⟨23⟩⟨25⟩[25][45]}+\\
\small\frac{[35][12]⟨45⟩^2[34]}{⟨12⟩⟨25⟩[25][45]⟨5|1\tiny{+}2|5]}\phantom{+}
\end{gathered}\nonumber
\end{equation}
\\[-92mm]
\begin{equation}\begin{gathered}
\small \tilde{r}^{--}_{66} = \frac{[35][12]⟨45⟩^2[34]}{⟨12⟩[45]⟨5|1\tiny{+}2|5]^2}
\end{gathered}\nonumber
\end{equation}
&
\begin{equation}\hspace{-3mm}\begin{gathered}
\small \tilde{r}^{--}_{85} = \frac{[23]⟨45⟩^4⟨12⟩⟨23⟩^2}{⟨13⟩^3⟨24⟩^2⟨25⟩^2⟨2|1\tiny{+}3|2]}
\end{gathered}\hspace{-2mm}\nonumber
\end{equation}
&
\vspace{80mm}
\begin{equation}\hspace{-2mm}\begin{gathered}
\small \tilde{r}^{--}_{100} = \frac{[35]⟨25⟩^2⟨15⟩[25]⟨45⟩^2}{⟨12⟩^3⟨35⟩⟨5|1\tiny{+}2|5]^2}+\\
\small\frac{-3[35]⟨25⟩^2⟨15⟩⟨14⟩⟨45⟩}{⟨12⟩^4⟨35⟩⟨5|1\tiny{+}2|5]}\phantom{+}
\end{gathered}\hspace{-5mm}\nonumber
\end{equation}
\\[-103mm]
\begin{equation}\begin{gathered}
\small \tilde{r}^{--}_{67} = \frac{⟨14⟩^2[24]⟨24⟩^2[34]}{⟨12⟩^4[25]⟨34⟩[45]}
\end{gathered}\nonumber
\end{equation}
&
\begin{equation}\begin{gathered}
\small \tilde{r}^{--}_{86} = \frac{⟨24⟩⟨25⟩[13][23]⟨35⟩}{⟨12⟩⟨23⟩^2[34]⟨2|1\tiny{+}5|2]}
\end{gathered}\nonumber
\end{equation}
&
\vspace{93mm}
\begin{equation}\hspace{-2mm}\begin{gathered}
\small \tilde{r}^{--}_{101} = \frac{[34]⟨34⟩⟨45⟩^2[15][14]}{⟨23⟩^2[45]⟨4|1\tiny{+}5|4]^2}+\\
\small\frac{-⟨24⟩[23]⟨34⟩⟨45⟩[14]^2}{⟨23⟩^2[45]⟨4|1\tiny{+}5|4]^2}+\\
\small(12345\; \rightarrow \; -32154)\phantom{+}
\end{gathered}\hspace{-5mm}\nonumber
\end{equation}
\\[-122mm]
\begin{equation}\begin{gathered}
\small \tilde{r}^{--}_{68} = \frac{⟨25⟩^3⟨24⟩⟨13⟩[25]}{⟨12⟩^3⟨23⟩^2⟨35⟩[45]}
\end{gathered}\nonumber
\end{equation}
&
\begin{equation}\begin{gathered}
\small \tilde{r}^{--}_{87} = \frac{⟨15⟩^2[23][25]⟨35⟩^2}{⟨13⟩^4[24]⟨25⟩[34]}+\\
\small(12345\; \rightarrow \; 12354)\phantom{+}
\end{gathered}\nonumber
\end{equation}
&
\\[-13mm]
\begin{equation}\hspace{-3mm}\begin{gathered}
\small \tilde{r}^{--}_{69} = \frac{[35]^2⟨35⟩^3[25]}{⟨13⟩⟨15⟩⟨23⟩⟨34⟩[45]^3}
\end{gathered}\hspace{-2mm}\nonumber
\end{equation}
&
\vspace{1mm}
\begin{equation}\begin{gathered}
\small \tilde{r}^{--}_{88} = \frac{[15]^2⟨45⟩^2⟨15⟩[14]}{⟨23⟩^2[45]⟨4|1\tiny{+}5|4]^2}+\\
\small(12345\; \rightarrow \; -32154)\phantom{+}
\end{gathered}\nonumber
\end{equation}
&
\\[-18mm]
\begin{equation}\hspace{-3mm}\begin{gathered}
\small \tilde{r}^{--}_{70} = \frac{[35]^2⟨45⟩⟨35⟩^2}{⟨12⟩⟨15⟩⟨23⟩⟨34⟩[45]^2}
\end{gathered}\hspace{-2mm}\nonumber
\end{equation}
&
\vspace{4mm}
\begin{equation}\begin{gathered}
\small \tilde{r}^{--}_{89} = \frac{⟨14⟩^2[13]^2⟨24⟩[14]}{⟨12⟩^3⟨15⟩[15]^3}+\\
\small\frac{⟨14⟩[13]^3⟨34⟩}{⟨12⟩^2⟨15⟩[15]^3}\phantom{+}
\end{gathered}\nonumber
\end{equation}
&
\\[-25mm]
\begin{equation}\hspace{-2mm}\begin{gathered}
\small \tilde{r}^{--}_{71} = \frac{⟨25⟩[13]^3⟨35⟩^2}{⟨12⟩[14]^3⟨15⟩⟨23⟩⟨24⟩}
\end{gathered}\hspace{-2mm}\nonumber
\end{equation}
&
\vspace{13mm}
\begin{equation}\begin{gathered}
\small \tilde{r}^{--}_{90} = \frac{-3⟨45⟩[12][13]}{⟨23⟩[45]⟨4|1\tiny{+}5|4]}+\\
\small\frac{[23][12]⟨45⟩^2}{⟨12⟩[24]⟨34⟩⟨4|1\tiny{+}5|4]}\phantom{+}
\end{gathered}\nonumber
\end{equation}
&
\\[-35mm]
\begin{equation}\begin{gathered}
\small \tilde{r}^{--}_{72} = \frac{⟨45⟩^3[14]⟨34⟩[24]}{⟨14⟩⟨24⟩⟨35⟩^3[45]^2}
\end{gathered}\nonumber
\end{equation}
&
\vspace{22mm}
\begin{equation}\begin{gathered}
\small \tilde{r}^{--}_{91} = \frac{2⟨24⟩^2[13][23]^2}{⟨12⟩⟨23⟩⟨35⟩[35]^3}+\\
\small\frac{⟨24⟩[13][23]⟨45⟩}{⟨12⟩⟨23⟩⟨35⟩[35]^2}\phantom{+}
\end{gathered}\nonumber
\end{equation}
&
\\[-44mm]
\begin{equation}\hspace{-3mm}\begin{gathered}
\small \tilde{r}^{--}_{73} = \frac{[14]^2[23]⟨14⟩^2}{⟨12⟩⟨13⟩[45]^2⟨1|2\tiny{+}3|1]}
\end{gathered}\hspace{-2mm}\nonumber
\end{equation}
&
&
\\[-8.5mm]
\begin{equation}\hspace{-3mm}\begin{gathered}
\small \tilde{r}^{--}_{74} = \frac{[12][23]^3⟨35⟩^2}{⟨12⟩[24]^3⟨34⟩⟨2|1\tiny{+}5|2]}
\end{gathered}\hspace{-2mm}\nonumber
\end{equation}
&
&
\\[-8.5mm]
\begin{equation}\hspace{-3mm}\begin{gathered}
\small \tilde{r}^{--}_{75} = \frac{⟨25⟩^2[23]^2⟨24⟩}{⟨12⟩^2⟨23⟩[34]⟨2|1\tiny{+}5|2]}
\end{gathered}\hspace{-2mm}\nonumber
\end{equation}
&
&
\\[-8.5mm]
\begin{equation}\hspace{-3mm}\begin{gathered}
\small \tilde{r}^{--}_{76} = \frac{[12]⟨15⟩⟨14⟩^2⟨25⟩}{⟨12⟩^2⟨13⟩^2⟨1|2\tiny{+}5|1]}
\end{gathered}\hspace{-2mm}\nonumber
\end{equation}
&
&
\end{tabular}

\pagebreak

\hskip-0.7cm
\begin{tabular}{p{5.4cm}p{5.8cm}p{5.9cm}}
\vspace{-6mm}
\begin{equation}\hspace{-3mm}\begin{gathered}
\small \tilde{r}^{--}_{102} = \frac{-2⟨14⟩[25]⟨35⟩^2⟨34⟩}{⟨13⟩^4[15]⟨23⟩}+\\
\small\frac{⟨34⟩⟨24⟩⟨35⟩^2[25]^2}{⟨13⟩^4[15]^2⟨23⟩}+\\
\small\frac{-2[12]⟨35⟩⟨34⟩^2[35]}{⟨13⟩^3[15]^2⟨23⟩}\phantom{+}
\end{gathered}\hspace{-2mm}\nonumber
\end{equation}
&
\vspace{-6mm}
\begin{equation}\hspace{-2mm}\begin{gathered}
\small \tilde{r}^{--}_{108} = \frac{⟨45⟩^3[13]^2⟨13⟩^2}{⟨12⟩^2⟨34⟩⟨35⟩⟨3|1\tiny{+}2|3]^2}+\\
\small\frac{-⟨24⟩⟨13⟩^2[13]⟨45⟩^3}{⟨12⟩^3⟨34⟩^2⟨35⟩⟨3|1\tiny{+}2|3]}+\\
\small\frac{-⟨45⟩^3[13]⟨25⟩⟨13⟩^2}{⟨12⟩^3⟨34⟩⟨35⟩^2⟨3|1\tiny{+}2|3]}\phantom{+}
\end{gathered}\hspace{-6mm}\nonumber
\end{equation}
&
\vspace{-6mm}
\begin{equation}\begin{gathered}
\small \tilde{r}^{--}_{114} = \frac{[12]^2[23]^2}{⟨13⟩[15][24]^2[35]}+\\
\small\frac{⟨45⟩^3⟨2|1\tiny{+}5|2]}{⟨13⟩⟨15⟩⟨23⟩⟨24⟩^2[24]}+\\
\small\frac{-[12]⟨45⟩^2⟨25⟩[25]}{⟨13⟩⟨23⟩⟨24⟩^2[24]^2}+\\
\small\frac{[23]^2⟨45⟩^2⟨5|1\tiny{+}3|5]}{⟨13⟩⟨15⟩⟨24⟩^2[24]^2[35]}+\\
\small\frac{-[12][23]^2[13]⟨34⟩}{⟨13⟩[15][24][35]⟨3|1\tiny{+}5|3]}+\\
\small\frac{-2[12]^2[23]^2[13]}{[15][24]^2[35]⟨3|1\tiny{+}5|3]}+\\
\small\frac{-[23]^2[13]⟨45⟩^2[15]}{⟨24⟩^2[24]^2[35]⟨3|1\tiny{+}5|3]}+\\
\small\frac{-2[23]^3⟨34⟩⟨35⟩}{⟨13⟩^2[24][35]⟨3|1\tiny{+}5|3]}+\\
\small\frac{[23]^2⟨35⟩⟨34⟩⟨14⟩[13]^2}{⟨13⟩⟨24⟩[24][35]⟨3|1\tiny{+}5|3]^2}+\\
\small\frac{3[12][23]^3[13]⟨35⟩}{[24]^2[35]⟨3|1\tiny{+}5|3]^2}+\\
\small\frac{3[23]^3[13]^2⟨34⟩⟨35⟩}{[24][35]⟨3|1\tiny{+}5|3]^3}+\\
\small\frac{-3[23]^3[13]⟨34⟩⟨35⟩^2}{⟨13⟩[24]⟨3|1\tiny{+}5|3]^3}\phantom{+}
\end{gathered}\nonumber
\end{equation}
\\[-100mm]
\begin{equation}\hspace{-4mm}\begin{gathered}
\small \tilde{r}^{--}_{103} = \frac{3⟨45⟩⟨35⟩^2[25]^2⟨15⟩^2}{⟨13⟩^4[24]⟨25⟩⟨5|1\tiny{+}3|5]}+\\
\small\frac{3[23]⟨35⟩^2[25]⟨15⟩^2}{⟨13⟩^4[24]⟨25⟩[34]}+\\
\small\frac{-⟨34⟩^2⟨45⟩[24]}{⟨13⟩^3[15]⟨23⟩}\phantom{+}
\end{gathered}\hspace{-6mm}\nonumber
\end{equation}
&
\begin{equation}\hspace{-2mm}\begin{gathered}
\small \tilde{r}^{--}_{109} = \frac{-⟨34⟩⟨35⟩[13][23]⟨45⟩}{⟨13⟩⟨23⟩⟨3|1\tiny{+}2|3]^2}+\\
\small\frac{-2⟨34⟩[23]^2⟨35⟩^2⟨24⟩}{⟨13⟩^2⟨23⟩⟨3|1\tiny{+}2|3]^2}+\\
\small\frac{2⟨34⟩[23]⟨35⟩^2⟨24⟩}{⟨13⟩^2⟨23⟩^2⟨3|1\tiny{+}2|3]}+\\
\small(12345\; \rightarrow \; -21354)\phantom{+}
\end{gathered}\hspace{-5mm}\nonumber
\end{equation}
&
\vspace{90mm}
\begin{equation}\hspace{-1mm}\begin{gathered}
\small \tilde{r}^{--}_{115} = \frac{-2/3[24]^2⟨24⟩⟨45⟩⟨14⟩^2}{⟨12⟩⟨13⟩^3⟨35⟩[35]^2}+\\
\small\frac{2/3⟨24⟩^2[23][24]^2[25]⟨14⟩}{⟨13⟩^3⟨35⟩[35]^3[45]}+\\
\small\frac{-[23]^3[13]⟨5|1\tiny{+}3|5]}{⟨13⟩[24][34]⟨35⟩[35]^3}+\\
\small\frac{5/3⟨24⟩[23][24]⟨14⟩^2[13]}{⟨12⟩⟨13⟩^2⟨35⟩[35]^3}+\\
\small\frac{-5/6[23]⟨2|1\tiny{+}3|2]⟨45⟩⟨14⟩}{⟨12⟩⟨13⟩^2⟨35⟩[35]^2}+\\
\small\frac{2/3[23][13]⟨14⟩⟨2|5\tiny{-}4|2]}{⟨12⟩⟨13⟩⟨35⟩[35]^2[45]}+\\
\small\frac{1/6⟨25⟩^2[23][25]^2⟨14⟩}{⟨12⟩⟨13⟩^2⟨35⟩[35]^2[45]}+\\
\small\frac{[23]⟨25⟩[25]⟨14⟩⟨45⟩}{⟨12⟩⟨13⟩^2⟨35⟩[35]^2}+\\
\small\frac{-1/2⟨25⟩⟨45⟩[25]^2⟨15⟩}{⟨12⟩⟨13⟩^2⟨35⟩[35][45]}+\\
\small\frac{-1/2⟨25⟩[23]^2⟨5|1\tiny{+}3|5]}{⟨12⟩⟨13⟩[34]⟨35⟩[35][45]}\phantom{+}
\end{gathered}\hspace{-6mm}\nonumber
\end{equation}
\\[-173mm]
\begin{equation}\hspace{-4mm}\begin{gathered}
\small \tilde{r}^{--}_{104} = \frac{2⟨24⟩^2⟨34⟩^2⟨5|3\tiny{-}4|2]}{⟨14⟩^2⟨23⟩^4[25]}+\\
\small\frac{-⟨12⟩[12]^2⟨34⟩^2⟨24⟩}{⟨14⟩⟨23⟩^4[25]^2}+\\
\small\frac{-2[12]⟨24⟩⟨34⟩^2[23]}{⟨14⟩⟨23⟩^3[25]^2}\phantom{+}
\end{gathered}\hspace{-5mm}\nonumber
\end{equation}
&
\vspace{1mm}
\begin{equation}\hspace{-2mm}\begin{gathered}
\small \tilde{r}^{--}_{110} = \frac{-2⟨25⟩⟨45⟩[25][13]⟨35⟩}{⟨13⟩[14]⟨15⟩⟨23⟩⟨24⟩[45]}+\\
\small\frac{[12]⟨45⟩[23][34]}{⟨15⟩[45]^2⟨4|1\tiny{+}2|4]}+\\
\small\frac{[35]⟨23⟩⟨45⟩^2[24]^2[13]}{⟨13⟩[14]⟨15⟩⟨24⟩[45]^2⟨4|1\tiny{+}2|4]}\phantom{+}
\end{gathered}\hspace{-10mm}\nonumber
\end{equation}
&
\\[-8.5mm]
\begin{equation}\hspace{-6mm}\begin{gathered}
\small \tilde{r}^{--}_{105} = \frac{-2⟨12⟩[12]^3[13]}{⟨13⟩[14][15][45]⟨1|2\tiny{+}3|1]}+\\
\small\frac{-[12]⟨24⟩[34]⟨34⟩^2}{⟨13⟩⟨14⟩[15]⟨23⟩^2[45]}+\\
\small\frac{[35][13]⟨35⟩^3}{⟨13⟩[14]⟨15⟩⟨23⟩^2[45]}\phantom{+}
\end{gathered}\hspace{-6mm}\nonumber
\end{equation}
&
\begin{equation}\begin{gathered}
\small \tilde{r}^{--}_{111} = \frac{[35]⟨15⟩⟨45⟩⟨35⟩}{⟨12⟩^2⟨13⟩⟨23⟩[24]}+\\
\small\frac{⟨25⟩⟨15⟩⟨14⟩[45]⟨45⟩}{⟨12⟩^3⟨13⟩⟨23⟩[24]}+\\
\small\frac{-3[35]⟨25⟩⟨15⟩⟨14⟩⟨35⟩}{⟨12⟩^3⟨13⟩⟨23⟩[24]}+\\
\small\frac{-3[35]⟨25⟩^2⟨14⟩[34]⟨35⟩}{⟨12⟩^4[14]⟨23⟩[24]}\phantom{+}
\end{gathered}\nonumber
\end{equation}
&
\\[-15mm]
\begin{equation}\hspace{-3mm}\begin{gathered}
\small \tilde{r}^{--}_{106} = \frac{2⟨15⟩⟨45⟩[13][23]}{⟨12⟩⟨14⟩[34]⟨35⟩[45]}+\\
\small\frac{2⟨25⟩⟨45⟩[24][13]}{⟨12⟩⟨23⟩⟨35⟩[34][45]}+\\
\small\frac{-[35]⟨25⟩⟨45⟩^2[23]}{⟨12⟩⟨14⟩⟨24⟩[34]⟨35⟩[45]}\phantom{+}
\end{gathered}\hspace{-5mm}\nonumber
\end{equation}
&
\vspace{5mm}
\begin{equation}\hspace{-3mm}\begin{gathered}
\small \tilde{r}^{--}_{112} = \frac{-3⟨14⟩[12]^2⟨15⟩^2[13]^2⟨34⟩}{⟨13⟩⟨1|2\tiny{+}5|1]^4}+\\
\small\frac{2⟨14⟩[12]⟨15⟩^2[13][15]⟨45⟩}{⟨12⟩⟨13⟩⟨1|2\tiny{+}5|1]^3}+\\
\small\frac{-2⟨14⟩[12]^2⟨15⟩^2[13]⟨34⟩}{⟨13⟩^2⟨1|2\tiny{+}5|1]^3}+\\
\small\frac{-[12]^2⟨15⟩⟨14⟩^2⟨35⟩}{⟨13⟩^3⟨1|2\tiny{+}5|1]^2}\phantom{+}
\end{gathered}\hspace{-8mm}\nonumber
\end{equation}
&
\\[-28mm]
\begin{equation}\hspace{-3mm}\begin{gathered}
\small \tilde{r}^{--}_{107} = \frac{-[25]⟨25⟩⟨15⟩^2⟨14⟩[13]}{⟨12⟩^2⟨13⟩^2⟨34⟩[34]^2}+\\
\small\frac{-[24]⟨14⟩^2⟨24⟩[13]}{⟨12⟩^2⟨13⟩⟨34⟩[35][45]}+\\
\small\frac{2[13]⟨14⟩^2⟨25⟩[23]}{⟨12⟩^2⟨13⟩⟨34⟩[34][35]}\phantom{+}
\end{gathered}\hspace{-8mm}\nonumber
\end{equation}
&
\vspace{17mm}
\begin{equation}\hspace{-5mm}\begin{gathered}
\small \tilde{r}^{--}_{113} = \frac{2[35]⟨34⟩⟨5|1\tiny{+}2|5]⟨14⟩^2[34]⟨15⟩}{⟨12⟩^4⟨13⟩⟨25⟩[25]^3}+\\
\small\frac{2⟨34⟩⟨5|1\tiny{+}2|5]⟨14⟩^2[34][23]}{⟨12⟩^3⟨13⟩⟨25⟩[25]^3}+\\
\small\frac{2⟨34⟩⟨14⟩^2[34][12][23]}{⟨12⟩^2⟨13⟩⟨25⟩[25]^3}+\\
\small\frac{⟨45⟩⟨14⟩^2⟨34⟩[34]^2}{⟨12⟩^3⟨13⟩⟨25⟩[25]^2}\phantom{+}
\end{gathered}\hspace{-14mm}\nonumber
\end{equation}
\end{tabular}

\newgeometry{left=1in, right=1in, top=1in, bottom=1in}  

\section{Reference evaluations}
\label{sec:referenceEvaluations}

To facilitate comparison with our results, we provide a numerical evaluation of the hard function defined in \cref{eq:hard-function,eq:hard-function-partial}.
We evaluate $B$ and $H^{(L),(n_c,n_f)}$ on the phase-space point
\begin{equation}
  \begin{aligned}
    p_1 &= \{-3.6033749869055013, 3.5549594933215615, 0.033937560795432568, 0.58772658529828721\},\\
    p_2 &= \{-3.5779991067160259, 3.5333697062718894, 0.033731453043168395, -0.56235070510881185\},\\
    p_3 &= \{1.7967619455543639, -1.7454551820128482, 0.10779867188908804, -0.41245501926361226\},\\
    p_4 &= \{0.41554983516150722, -0.38259279840807596, -0.13137170294301400, 0.095109893149375643\},\\
    p_5 &= \{4.9690623129056561, -4.9602812191725268, -0.044095982784675001, 0.29196924592476125\},
  \end{aligned}
\end{equation}
and the renormalization scale is set to $\mu=1$.

The reference evaluations are
\begin{equation}
  \begin{tabular}{MMM}
    B  & = &154023.6666921499, \\[2em]
    H^{(1),(1,0)}  &= & -1.616272307398762, \\
    H^{(1),(-1,0)} & = & 1.410723596231933, \\[1em]
    H^{(1),(0,1)} & = & 2.508992268209689, \\
    H^{(1),(-2,1)} & = & -0.04224997482253675,
  \end{tabular}
  \qquad \qquad
  \begin{tabular}{MMM}
    H^{(2),(2,0)} &  =&  39.18504944322860, \\
    H^{(2),(0,0)} & = & 181.8421831216970, \\[1em]
    H^{(2),(1,1)} & = &  -3.524523895322017, \\
    H^{(2),(-1,1)}&  =&   -6.578862105257153, \\
    H^{(2),(-3,1)}&  =&   0.01653881075908589, \\[1em]
    H^{(2),(0,2)} & = &  1.703372292225135, \\
    H^{(2),(-2,2)}&  =&   -0.4017699546907486, \\
    H^{(2),(-4,2)}&  =&   1.039350843363634.
  \end{tabular}
\end{equation}

\twocolumngrid
\bibliography{main_5g}

\end{document}